\documentclass[conference]{IEEEtran}
\usepackage[left=0.67in,right=0.67in,top=0.64in,bottom=0.70in]{geometry}
\usepackage[T1]{fontenc}
\usepackage[latin9]{inputenc}
\usepackage{amsthm}
\usepackage{amsmath} 
\usepackage{amsfonts}
\usepackage{graphicx} 
\usepackage{color} 
\usepackage{cite}
\usepackage{algpseudocode}
\usepackage{algorithm}
\usepackage{amssymb}
\usepackage{subfigure}
\usepackage{esint}
\usepackage{algpseudocode}
\usepackage{epstopdf}
\usepackage{multirow}
\usepackage{makecell}
\usepackage{float}
\usepackage{lipsum}
\usepackage{balance}
\usepackage{color}
\usepackage{hyperref}
\usepackage{tikz}
\usepackage{bbm}
\usetikzlibrary{shapes ,arrows, positioning} 

\newtheorem{remark}{Remark}

\newtheorem{example}{Example}

\theoremstyle{plain}
\theoremstyle{plain}
\newtheorem{theorem}{Theorem}
\newtheorem{lemma}{Lemma}

\newcommand{\comment}[1]{}

\IEEEoverridecommandlockouts

\allowdisplaybreaks

\algnewcommand\algorithmicforeach{\textbf{for each}}
\algdef{S}[FOR]{ForEach}[1]{\algorithmicforeach\ #1\ \algorithmicdo}

\definecolor{ogreen}{rgb}{0,0.6,0}

\begin{document}

\title{Adapt or Regress: Rate-Memory-Compatible Spatially-Coupled Codes\vspace{-0.4em}}

\author{
   \IEEEauthorblockN{Bade Aksoy, Do\u{g}ukan \"{O}zbayrak, and Ahmed Hareedy}
   \IEEEauthorblockA{Department of Electrical and Electronics Engineering, Middle East Technical University, 06800 Ankara, Turkey \\ bade.aksoy@metu.edu.tr, dogukan.ozbayrak@metu.edu.tr, and ahareedy@metu.edu.tr\vspace{-1.3em}}
}
\maketitle

\begin{abstract}

From wireless communications to data storage, low-density parity-check (LDPC) codes are being increasingly employed to fortify data in modern systems. Spatially-coupled (SC) codes are a class of LDPC codes that have excellent performance in both the asymptotic and the finite-length regimes thanks to the degrees of freedom they offer. An SC code is designed by partitioning a base matrix into components, the number of which implies the code memory, then coupling and lifting them. In the same system, various error-correction coding schemes are typically needed. For example, in wireless communication standards, several channel conditions and data rates should be supported. In storage and computing systems, stronger codes should be adopted as the device ages. Adaptive code design enables switching from one code to another when needed, ensuring reliability while reducing hardware cost. In this paper, we introduce a class of reconfigurable SC codes named rate-memory-compatible SC (RMC-SC) codes, which we design probabilistically. In particular, rate compatibility in RMC-SC codes is achieved via increasing the SC code memory, which also makes the codes memory-compatible and improves performance. We express the expected number of short cycles in the SC code protograph as a function of the fixed probability distribution characterizing the already-designed SC code as well as the unknown distribution characterizing the additional components. We use the gradient-descent algorithm to find a locally-optimal distribution, in terms of cycle count, for the new components. The method can be recursively used to design any number of SC codes needed, and we show how to extend it to the case when rate compatibility is achieved via adding rows (or columns) instead. Next, we perform the finite-length optimization using a Markov chain Monte Carlo (MC\textsuperscript{2}) approach that we update to design the proposed RMC-SC codes. Experimental results demonstrate significant reductions in cycle counts as well as remarkable performance gains achieved by RMC-SC codes compared with a straightforward scheme. Based on the system requirements, machine learning can be used to reconfigure RMC-SC codes.

\end{abstract}

\begin{IEEEkeywords}
LDPC codes, rate-compatible codes, spatially-coupled codes, gradient descent, MCMC methods.
\end{IEEEkeywords}

\section{Introduction}\label{sec_intro}

Low-density parity-check (LDPC) codes \cite{gal_th} are widely employed as an effective error-correction technique in a variety of modern systems. They find application in wireless communications \cite{declercq} as well as in data storage technologies \cite{ahh_jsac}. Spatially-coupled (SC) codes are a class of LDPC codes that is increasingly gaining attention since it is proven to be capacity-approaching \cite{kudekar} and demonstrated to offer excellent finite-length performance \cite{costello}. Here, we are interested in time-invariant protograph-based SC codes, where the design has two phases. A protograph base matrix is first partitioned into components that are coupled to form the code protograph \cite{battag_sc, oocpo}. Next, the protograph matrix is lifted to reach the final parity-check matrix of the SC code \cite{oocpo, rosnes}. The number of component matrices is the SC code memory plus one, and~as this memory increases, the performance notably improves \cite{banihashemi}. Approaches based on discrete optimization were offered to design high-performance SC codes with low memories \cite{oocpo, channel_aware}. However, such approaches struggle, complexity-wise, when there are additional degrees of freedom in the SC code design. Recently, probabilistic approaches were presented to address this challenge and design effective SC and multi-dimensional SC codes with high memories \cite{ahh_grade, reins_gdmd}. Furthermore, a method for optimizing partitioning and lifting of SC codes based on Markov chain Monte Carlo algorithms was introduced in \cite{reins_mcmc} (see also \cite{mackay_it}).

Adaptive, rate-compatible (RC) codes enable the system to have multiple error-correction schemes, to be used based on the need, with remarkable savings in hardware \cite{chen_wesel}. In wireless communications, different codes should be used based on the channel type (data versus control) and condition as well as the transmission rate. In data storage, using the~same coding scheme designed for a fresh device late in its lifetime causes performance to notably regress. Raptor-like RC-LDPC codes as well as RC-LDPC codes designed algebraically for communication systems were introduced in \cite{chen_wesel} and \cite{battaglioni_rc}, respectively. RC Bose-Chaudhuri-Hocquenghem (BCH) and LDPC codes based on syndrome-coupling that can be used in Flash memories were presented in \cite{huang}. RC non-binary LDPC codes developed via optimized puncturing distributions were introduced in \cite{savin}. Targeting systems adopting hybrid automatic repeat request (HARQ), RC spatially-coupled (RC-SC) codes were designed in \cite{liu_rc} and \cite{yu_rc}. Other RC-SC code designs based on multi-edge type (MET) ensembles \cite{nitzold}, random puncturing \cite{mitchell}, and progressive-edge growth (PEG) methods \cite{bani_rc, he_rc} also exist in the literature.

In this paper, we present the first probabilistic framework to design high-performance SC codes that are rate-compatible and memory-compatible, which we name \textit{rate-memory-compatible SC (RMC-SC) codes}. We focus on binary codes. Redundancy increments are achieved via increasing the memory of the RMC-SC code, which naturally improves performance. Our codes are designed recursively on various design stages where we adopt three procedures. First, we express the expected number of cycles in the protograph of the RMC-SC code in terms of a fixed distribution, obtained from the code designed at the previous stage, and a new or an optimizable distribution, which guides the design of the new components. This expectation is our objective function to minimize, and we derive the gradients as well as the solution form, offering detailed expressions for cycles-$4$, cycles-$6$, and cycles-$8$. Next, we use a gradient-descent algorithm to find this new distribution that locally minimizes the objective function. Second, we adopt a Markov chain Monte Carlo (MCMC or MC\textsuperscript{2}) method, which takes the gradient-descent distribution as input, to specify the finite-length (FL) partitioning across the new components. Third, we employ this MC\textsuperscript{2} method on the level of lifting to design the final parity-check matrix of the RMC-SC code, further minimizing the multiplicity of short cycles. Moreover, we show how our framework can be extended to design RC-SC codes where redundancy control is implemented via adding rows (or columns). We present experimental results that show how much reduction in short cycle counts RMC-SC codes can achieve as well as how much performance gains they can offer (for the last design stage, the results are close) compared with RC codes designed via a straightforward approach that only optimizes the SC code with the highest memory based on recent literature.

The rest of the paper is organized as follows. In Section~\ref{sec_design}, we discuss preliminaries and introduce the RMC-SC code design idea. In Section~\ref{sec_prob}, we present the probabilistic analysis of our framework by deriving the required expectations. In Section~\ref{sec_grad}, we characterize the solution form and present our gradient-descent algorithm. In Section~\ref{sec_mcmc}, we introduce the MC\textsuperscript{2} FL methods to design the codes. In Section~\ref{sec_exp}, we offer the experimental results. In Section~\ref{sec_conc}, we conclude the~paper.

\vspace{-0.4em}
\section{RMC-SC Code Design Idea}\label{sec_design}

The protograph matrix of an SC code $\mathbf{H}_{\textup{SC}}$ is designed via partitioning a protograph base matrix into $m+1$ disjoint components $\mathbf{H}_0, \mathbf{H}_1, \dots, \mathbf{H}_m$ such that they sum to this matrix itself, where $m$ is defined as the memory of the code. Then, these components are coupled $L$ times, where $L$ is defined as the coupling length of the code. Equation \eqref{sc_matrix} depicts the protograph matrix of an SC code with $m=m_{\textup{f}}+m_{\textup{n}}$ (these two will be defined shortly). Next, lifting is performed, where each $1$ (resp., $0$) in the protograph matrix is replaced by a $z\times z$ circulant (resp., all-zero matrix), where $z$ is defined as a lifting parameter, to design the parity-check matrix of the SC code $\mathbf{H}_{\textup{SC}}^{\textup{C}}$. A circulant is an identity matrix with its columns shifted one unit cyclically to the left that is raised to some power $t$ in $\{0,1,\dots,z-1\}$. A collection of columns with the components aligned vertically is called a replica \cite{channel_aware}. Next, we discuss the core idea of the RMC-SC code design.

\vspace{-1.0em}\begin{gather} \label{sc_matrix}
\hspace{-0.3em}\mathbf{H}_{\textup{SC}} =
\hspace{-0.3em}\begin{bmatrix}
\textcolor{blue}{\mathbf{H}_0} & \mathbf{0}  &  & & & \mathbf{0} \vspace{-0.5em}\\
\textcolor{blue}{\vdots} & \textcolor{blue}{\mathbf{H}_0} & &  &  & \vdots \vspace{-0.5em}\\
\textcolor{blue}{\mathbf{H}_{m_{\textup{f}}}} & \textcolor{blue}{\vdots} & \textcolor{blue}{\ddots} &  &  & \vdots \vspace{-0.5em}\\
\textcolor{ogreen}{\mathbf{H}_{m_{\textup{f}}+1}} & \textcolor{blue}{\mathbf{H}_{m_{\textup{f}}}} & \textcolor{blue}{\ddots} & \textcolor{blue}{\ddots} & & \vdots \vspace{-0.5em}\\
\textcolor{ogreen}{\vdots} & \textcolor{ogreen}{\mathbf{H}_{m_{\textup{f}}+1}} & \textcolor{blue}{\ddots} & \textcolor{blue}{\ddots} & \textcolor{blue}{\ddots}  & \mathbf{0} \vspace{-0.5em}\\
\textcolor{ogreen}{\mathbf{H}_{m_{\textup{f}}+m_{\textup{n}}}} & \textcolor{ogreen}{\vdots} & \textcolor{ogreen}{\ddots} & \textcolor{blue}{\ddots} & \textcolor{blue}{\ddots} & \textcolor{blue}{\mathbf{H}_0} \vspace{-0.5em}\\
\mathbf{0} & \textcolor{ogreen}{\mathbf{H}_{m_{\textup{f}}+m_{\textup{n}}}} & \textcolor{ogreen}{\ddots} & \textcolor{ogreen}{\ddots} & \textcolor{blue}{\ddots} & \textcolor{blue}{\vdots} \vspace{-0.5em}\\
\vdots & \mathbf{0}& \textcolor{ogreen}{\ddots} & \textcolor{ogreen}{\ddots} & \textcolor{ogreen}{\ddots} & \textcolor{blue}{\mathbf{H}_{m_{\textup{f}}}}\vspace{-0.5em}\\
\vdots & \vdots& & \textcolor{ogreen}{\ddots} & \textcolor{ogreen}{\ddots} & \textcolor{ogreen}{\mathbf{H}_{m_{\textup{f}}+1}}\vspace{-0.5em}\\
\vdots & \vdots& &  & \textcolor{ogreen}{\ddots} & \textcolor{ogreen}{\vdots}\vspace{-0.0em}\\
\mathbf{0} & \mathbf{0}& &  &  & \textcolor{ogreen}{\mathbf{H}_{m_{\textup{f}}+m_{\textup{n}}}}
\end{bmatrix}.
\end{gather}

Short cycles degrade the performance of LDPC codes, both in the waterfall region, by creating dependencies within message-passing decoders, and in the error-floor region, by being common substructures in more detrimental objects such as absorbing sets \cite{oocpo, banihashemi, channel_aware}. A cycle-$2\ell$ candidate is a way of traversing a protograph pattern to generate cycles-$2\ell$ after partitioning/lifting \cite{channel_aware}. We focus on patterns and cycle-$2\ell$ candidates that have $2\ell$ distinct entries because of their multiplicity dominance \cite{ahh_grade}. In this paper, whenever we discuss cycles in the protograph context, we mean cycle~candidates.

Suppose that the goal is to design $s+1$ RMC-SC codes, which means we have $s+1$ recursive design stages. We start with the protograph design of the codes. Initially, an all-one matrix $\mathbf{H}$ of dimension $\gamma \times \kappa$, $\gamma \geq 4$ and $\kappa > \gamma$, is partitioned into $s+1$ disjoint matrices $\mathbf{H}_{\textup{n},0}, \mathbf{H}_{\textup{n},1}, \dots, \mathbf{H}_{\textup{n},s}$, each of which has the same dimension, $\gamma \times \kappa$, where $\sum_{d=0}^s \mathbf{H}_{\textup{n},d} = \mathbf{H}$. We denote the probability that a $1$ in $\mathbf{H}$ is assigned to $\mathbf{H}_{\textup{n},d}$ by $r_{\textup{n},d} > 0$, for all $d$ in $\{0,1,\dots,s\}$, where $\sum_{d=0}^s r_{\textup{n},d} = 1$. For Design Stage~$d$, we introduce the matrix $\mathbf{H}_{\textup{f},d} = \sum_{j=0}^{d-1} \mathbf{H}_{\textup{n},j}$ and the probability $r_{\textup{f},d} = \sum_{j=0}^{d-1} r_{\textup{n},j}$, for all $d$ in $\{1,2,\dots,s\}$, leading to the recursive relations:
\begin{align}
\hspace{-0.5em}\mathbf{H}_{\textup{f},d} &= \mathbf{H}_{\textup{f},d-1} + \mathbf{H}_{\textup{n},d-1}, \textup{ } d \in \{2,3,\dots,s\}, \textup{ } \mathbf{H}_{\textup{f},1}=\mathbf{H}_{\textup{n},0}. \nonumber \\
r_{\textup{f},d} &= r_{\textup{f},d-1} + r_{\textup{n},d-1}, \textup{ } d \in \{2,3,\dots,s\}, \textup{ } r_{\textup{f},1}=r_{\textup{n},0}. \nonumber
\end{align}
Here, the ``f'' in the subscripts is for ``fixed,'' while the ``n'' in the subscripts is for ``new'' throughout the paper. Moreover, at Design Stage~$d$, for all $d$ in $\{1,2,\dots,s\}$, the matrix $\mathbf{H}_{\textup{n},d}$ is partitioned into $m_{\textup{n},d} > 0$ disjoint components $\mathbf{H}_k, \textup{ } k = m_{\textup{f},d}+1, {m_{\textup{f},d}+2}, \dots, m_{\textup{f},d}+m_{\textup{n},d}$, where $\sum_{k=m_{\textup{f},d}+1}^{m_{\textup{f},d}+m_{\textup{n},d}} \mathbf{H}_k = \mathbf{H}_{\textup{n},d}$. As for Design Stage~0, the matrix $\mathbf{H}_{\textup{n},0}$ is partitioned into $m_{\textup{n},0}+1$ disjoint components $\mathbf{H}_k, \textup{ } k = 0, 1, \dots, m_{\textup{n},0}$, where $\sum_{k=0}^{m_{\textup{n},0}} \mathbf{H}_k = \mathbf{H}_{\textup{n},0}$. Additionally, we have $\sum_{d=0}^s m_{\textup{n},d} = m_s$. Here and once again, for Design Stage~$d$, we define the memory $m_{\textup{f},d} = \sum_{j=0}^{d-1} m_{\textup{n},j}$, for all $d$ in $\{1,2,\dots,s\}$, leading to the recursive relation:
\begin{equation}
m_{\textup{f},d} = m_{\textup{f},d-1} + m_{\textup{n},d-1}, \textup{ } d \in \{2,3,\dots,s\}, \textup{ } m_{\textup{f},1}=m_{\textup{n},0}. \nonumber
\end{equation}

The idea of our RMC-SC code design is that at Design Stage~$d$, $1 \leq d \leq s$, there is a fixed protograph base matrix $\mathbf{H}_{\textup{f},d}$, with parameters $r_{\textup{f},d}$ and $m_{\textup{f},d}$ along with its partitioning, inherited from Design Stage~$d-1$ as it characterizes RMC-SC Code~$d-1$. The objective of Design Stage~$d$ is to effectively (near-optimally with respect to cycle count) partition a new matrix $\mathbf{H}_{\textup{n},d}$, with parameters $r_{\textup{n},d}$ and $m_{\textup{n},d}$, in order to~complete the design of RMC-SC Code~$d$. Observe that RMC-SC Code~$d$ has protograph base matrix $\mathbf{H}_{\textup{f},d}+\mathbf{H}_{\textup{n},d}$, probability $r_{\textup{f},d}+r_{\textup{n},d}$, memory $m_{\textup{f},d}+m_{\textup{n},d}$, and $m_{\textup{n},d}$ additional component matrices compared with RMC-SC Code~$d-1$. Therefore, at each design stage, the code rate decreases (rate compatibility) as the memory increases (memory-compatibility). In the context of probabilistic analysis, the term ``fixed'' only means distribution-wise fixed, unlike in the finite-length context.

Design Stage~0 is simple, and the distribution of $1$'s in $\mathbf{H}_{\textup{n},0}$ across components is optimally obtained via gradient descent according to \cite{ahh_grade}. For notation simplicity, we drop the design stage index $d$ from the subscripts, for all $d$ in $\{1,2,\dots,s\}$, unless it is needed. Thus, we will typically discuss things in terms of $\mathbf{H}_{\textup{f}}$, with $r_{\textup{f}}$ and $m_{\textup{f}}$, on one side and $\mathbf{H}_{\textup{n}}$, with $r_{\textup{n}}$ and $m_{\textup{n}}$, on the other. We first discuss the probabilistic part of the code design for a generic design stage (not indexed by $0$). We denote the probability that a $1$ in $\mathbf{H}_{\textup{f}}$ is assigned to its component $\mathbf{H}_k$ by $p_k$, for all $k$ in $\{0,1,\dots,m_{\textup{f}}\}$. We denote the probability that a $1$ in $\mathbf{H}_{\textup{n}}$ is assigned to its component $\mathbf{H}_k$ by $q_k$, for all $k$ in $\{m_{\textup{f}}+1,m_{\textup{f}}+2,\dots,m_{\textup{f}}+m_{\textup{n}}\}$. Now, we have the two key probability, i.e., edge, distributions:
\vspace{-0.1em}\begin{equation}\label{eqn_dists}
\mathbf{p} = [p_0 \ p_1 \ \dots \ p_{m_{\textup{f}}}], \textup{ } \mathbf{q} = [q_{m_{\textup{f}}+1} \ q_{m_{\textup{f}}+2} \ \dots \ q_{m_{\textup{f}}+m_{\textup{n}}}].
\end{equation}
The distribution $\mathbf{p}$ is fixed since it characterizes the design of the RMC-SC code from the previous stage, while the distribution $\mathbf{q}$ is our variable of interest. These are the steps we follow to systematically find $\mathbf{q}$ at any design stage:
\begin{enumerate}
\item We express the expected number of cycles-$2\ell$ in the protograph of the RMC-SC code under random partitioning in terms of $r_{\textup{f}}$, $r_{\textup{n}}$, the entries of $\mathbf{p}$, and the entries of $\mathbf{q}$.
\item We develop a gradient-descent (GD) algorithm to find a locally-optimal $\mathbf{q}$ that minimizes such expectation (for cycles-$6$, cycles-$8$, or a combination).
\item The final distribution for the RMC-SC code is:
\begin{align}
\mathbf{u} = \left[ \frac{r_{\textup{f}}}{r_{\textup{f}}+r_{\textup{n}}}\mathbf{p} \textup{ }\textup{ }\textup{ } \frac{r_{\textup{n}}}{r_{\textup{f}}+r_{\textup{n}}}\mathbf{q} \right] \in [0,1]^{1\times(m_{\textup{f}}+m_{\textup{n}}+1)},
\end{align}
which will be the fixed distribution $\mathbf{p}$ in the next RMC-SC design stage.
\end{enumerate}

We note that in this paper, vectors are row vectors by default. Moreover, entries in protograph base matrices are assumed independent in the GD probabilistic analysis \cite{ahh_grade, reins_gdmd}. For the last code, i.e., at Design Stage~$s$, $\mathbf{H}_{\textup{f}}+\mathbf{H}_{\textup{n}} = \mathbf{H}$, $r_{\textup{f}}+r_{\textup{n}}=1$, and $m_{\textup{f}}+m_{\textup{n}}=m_s$. Therefore, the variable node (VN) degree or the column weight of this code is $\gamma$, and its maximum check node (CN) degree or its maximum row weight is $\kappa$.

Now, we introduce the finite-length (FL) design idea at any stage (not indexed by $0$), which is summarized as follows:
\begin{enumerate}
\item We develop an MC\textsuperscript{2} algorithm that takes the distribution $\mathbf{q}$ as input and finds an actual partitioning of $\mathbf{H}_{\textup{n}}$ that further minimizes the population of short cycles.
\item We extend this MC\textsuperscript{2} algorithm to also find the lifting parameters (circulant powers) of the $1$'s in $\mathbf{H}_{\textup{n}}$ that minimize the number of short cycles in the final parity-check matrix of the RMC-SC code $\mathbf{H}_{\textup{SC}}^{\textup{C}}$.
\end{enumerate}
Observe that here, $\mathbf{H}_{\textup{f}}$, its components, and their lifted versions are fixed matrices. Observe also that at any design stage other than the last, the RMC-SC code is very likely not to be variable-regular. More details, including FL design restrictions, will be discussed in the relevant sections.

The length of such RMC-SC code is $\kappa z L$, while its design rate is $1-\gamma(L+m_{\textup{f}}+m_{\textup{n}})/(\kappa L)$, which demonstrates how the rate is decreased as the memory is increased for a fixed length. The protograph matrix of this code $\mathbf{H}_{\textup{SC}}$ is given in \eqref{sc_matrix}, where the fixed part is in blue and the new part is in green.

We also define a partitioning matrix $\mathbf{K}$ and a lifting matrix $\mathbf{T}$, both of size $\gamma \times \kappa$. If $\mathbf{K}(i,j)=x \in \{0,1,\dots,m_{\textup{f}}+m_{\textup{n}}\}$ (resp., $\mathbf{T}(i,j)=t \in \{0,1,\dots,z-1\}$), this means entry $(i,j)$ in the protograph base matrix $\mathbf{H}_{\textup{f}}+\mathbf{H}_{\textup{n}}$ of the RMC-SC code goes to component matrix $\mathbf{H}_x$ (resp., is associated with a circulant of power $t$). If the corresponding entry in $\mathbf{H}$ is not assigned to $\mathbf{H}_{\textup{f}}+\mathbf{H}_{\textup{n}}$ of the relevant RMC-SC code, we set $x=t=-1$. At each stage, $\mathbf{K}$ and $\mathbf{T}$ also have fixed (non-negative $x \leq m_{\textup{f}}$) and optimizable ($x > m_{\textup{f}}$) entries.

\section{Probabilistic Analysis and Extension}\label{sec_prob}

In this section, we introduce the theoretical foundations of our probabilistic RMC-SC code design framework.

For an RMC-SC code, consider a cycle-$2\ell$ in the all-one protograph matrix $\mathbf{H}$ with entries $\{ (i_1,j_1), (i_1,j_2), (i_2,j_2), \allowbreak (i_2,j_3), \dots, (i_{\ell},j_{\ell}), (i_{\ell},j_{\ell+1}) \}$, where $j_{\ell+1}=j_1$. From \cite{battag_sc}, the condition that this cycle remains active after partitioning~is:
\begin{equation}\label{eqn_cond}
\sum_{k=1}^{\ell} \big[ \mathbf{K}(i_k,j_k)-\mathbf{K}(i_k,j_{k+1}) \big] = 0,
\end{equation}
assuming all of its entries are assigned to the protograph base matrix $\mathbf{H}_{\textup{f}}+\mathbf{H}_{\textup{n}}$ of the RMC-SC code.
This equation is inspired by the result in \cite{fossorier}. By replacing $\mathbf{K}$ with $\mathbf{T}$ and making the equality congruent $(\hspace{-0.6em}\mod z)$, we reach the equivalent equation for the case of lifting. After recalling \eqref{eqn_dists}, we define the following two polynomials
\vspace{-0.2em}\begin{equation}
f(X) = \sum_{j=0}^{m_{\textup{f}}} p_j X^j \textup{ and } g(X) = \hspace{-0.5em}\sum_{j=m_{\textup{f}}+1}^{m_{\textup{f}}+m_{\textup{n}}} q_j X^j,
\end{equation}
where $f(X)$ is associated with the components of $\mathbf{H}_{\textup{f}}$ while $g(X)$ is associated with the components of $\mathbf{H}_{\textup{n}}$. We define the notation $[h(X)]_i$, for any polynomial $h(X)$, as the coefficient of $X^i$ in this polynomial. We also note that all expectations here are $L$-invariant for simplicity.

In Theorem~\ref{thm:elegant objective function}, we express the expected number of cycle-$2\ell$ candidates in the RMC-SC protograph in terms of $r_{\textup{f}}$ and $r_{\textup{n}}$ as well as the entries of the distributions $\mathbf{p}$ and $\mathbf{q}$. Theorem~\ref{thm:elegant objective function} and Lemma~\ref{lemma: obj_fn} open the door for the gradient-descent expectation minimization to be discussed in Section~\ref{sec_grad}.

\begin{theorem} \label{thm:elegant objective function} 
Consider an RMC-SC code that has $\mathbf{H}_{\textup{f}}$, with $r_{\textup{f}}$ and $m_{\textup{f}}$, as well as $\mathbf{H}_{\textup{n}}$, with $r_{\textup{n}}$ and $m_{\textup{n}}$. Denote the expected number of cycle-$2\ell$ candidates that has $2\ell$ distinct entries in the RMC-SC protograph by $\mathbb{E}[\textup{cycle-$2\ell$}]$, where $\ell \in \mathbb{N}$ and $\ell \geq 2$. Then, 
\begin{align}\label{eqn_thm1}
& \mathbb{E}[\textup{cycle-}2\ell] = E_{2\ell}(\mathbf{q}) =  \nonumber \\
& A_{2\ell} \Big[ \big( r_{\textup{f}}f(X) +r_{\textup{n}}g(X) \big)^{\ell}  \big(r_{\textup{f}}f(X^{-1}) +r_{\textup{n}}g(X^{-1}) \big)^{\ell}        \Big]_{0}, 
\end{align}
where $A_{2\ell}$ is the number of all such cycle-$2\ell$ candidates in the all-one matrix $\mathbf{H}$.
\begin{proof}
We again consider a cycle-$2\ell$ candidate in $\mathbf{H}$ as described at the start of the section. Suppose that the RMC-SC code has partitioning matrix $\mathbf{K}$. Then, by linearity, $\mathbb{E}[\textup{cycle-}2\ell]$ can be expressed as follows:
\vspace{-0.1em}\begin{equation}\label{exp_cycle}
\mathbb{E}[\textup{cycle-}2\ell] = A_{2\ell}\ \mathbb{P}[\textup{cycle-}2\ell],
\end{equation}
where $\mathbb{P}[\textup{cycle-}2\ell]$ is the probability that a cycle-$2\ell$ candidate in $\mathbf{H}$ becomes a cycle-$2\ell$ candidate in the code protograph after partitioning. Define $z_{w}$ as the probability that an entry in $\mathbf{K}$ takes the value $w \in \{0,1,2, \dots, m_{\textup{f}}+m_{\textup{n}}\}$, and define $Z_w$ as the corresponding event. Moreover, define $p_w=0$, for all $w$ in $\{m_{\textup{f}}+1, m_{\textup{f}}+2, \dots, m_{\textup{f}}+m_{\textup{n}}\}$, and define $q_w=0$, for all $w$ in $\{0,1,\dots,m_{\textup{f}}\}$. Then,
\begin{align}
z_w &=\mathbb{P}[W_{\textup{f}}]\,\mathbb{P}[Z_{w}|W_{\textup{f}}] + \mathbb{P}[W_{\textup{n}}]\,\mathbb{P}[Z_{w}|W_{\textup{n}}]  \nonumber \\
&=r_{\textup{f}}\,p_{w} + r_{\textup{n}}\,q_{w}, 
\end{align}
where $W_{\textup{f}}$ (resp., $W_{\textup{n}}$) is the event that $w$ refers to one of the first $m_{\textup{f}}+1$ (resp., the last $m_{\textup{n}}$) components. Consider the condition in \eqref{eqn_cond} for a cycle-$2\ell$. Let $B$ be the event that all entries of the cycle-$2\ell$ candidate are assigned to the base matrix $\mathbf{H}_{\textup{f}}+\mathbf{H}_{\textup{n}}$ of the RMC-SC code. Moreover, let $x_k$ and $y_k$ be in $\{0,1,2, \dots, m_{\textup{f}}+m_{\textup{n}}\}$, for all $k$ in $\{1,2,\dots,\ell\}$. Then,
\vspace{-0.1em}\begin{align}\label{eqn_prob}
&\mathbb{P}[\textup{cycle-}2\ell] = P_{2\ell}(\mathbf{q}) \nonumber \\
&=\mathbb{P}\Big[ \Big( \sum_{k=1}^{\ell} \big[\mathbf{K}(i_k ,j_k)-\mathbf{K}(i_k ,j_{k+1}) \big] = 0 \Big) \cap B \Big]  \nonumber \\
&=\hspace{-1.3em}\sum_{\sum_{k=1}^{\ell}  (x_k-y_k)=0} \hspace{-0.6em}\Big( \prod_{k=1}^{\ell}\mathbb{P}\big[ \mathbf{K}(i_k,j_k)=x_k \big] \mathbb{P}\big[ \mathbf{K}(i_k,j_{k+1})=y_k \big] \Big) \nonumber \\
&=\hspace{-1.3em}\sum_{\sum_{k=1}^{\ell}  (x_k-y_k)=0} \hspace{+0.3em} \prod_{k=1}^{\ell} z_{x_k}z_{y_k} \nonumber \\
&=\hspace{-1.3em}\sum_{ \sum_{k=1}^{\ell}(x_k-y_k)=0} \hspace{+0.3em} \prod_{k=1}^{\ell}\big[ r_{\textup{f}}\,p_{x_k} + r_{\textup{n}}\,q_{x_k}    \big] \big[ r_{\textup{f}}\,p_{y_k} + r_{\textup{n}}\,q_{y_k}    \big] \nonumber \\
&=\hspace{-0.2em}\Big[\hspace{-2.0em}\sum_{x_k ,y_k \in \{0,1, \dots ,m_{\textup{f}}+m_{\textup{n}} \}, \forall k  } \hspace{+0.0em} \Big( \prod_{k=1}^{\ell} \big[ r_{\textup{f}}\,p_{x_k} + r_{\textup{n}}\,q_{x_k}    \big] \nonumber \\
& \hspace{10.0em} \cdot \big[ r_{\textup{f}}\,p_{y_k} + r_{\textup{n}}\,q_{y_k}    \big]  \Big) X^{\sum_{v=1}^{\ell}(x_v-y_v)} \Big]_0 \nonumber \\
&= \hspace{-0.2em}\Big[ \big( r_{\textup{f}}f(X) +r_{\textup{n}}g(X) \big)^{\ell}  \big(r_{\textup{f}}f(X^{-1}) +r_{\textup{n}}g(X^{-1}) \big)^{\ell}   \Big]_{0}.
\end{align}
Substituting \eqref{eqn_prob} in \eqref{exp_cycle} completes the proof.
\end{proof}

\end{theorem}

\begin{figure}
\centering
\includegraphics[trim={19.5in 3.0in 20.0in 3.0in},width=0.44\textwidth]{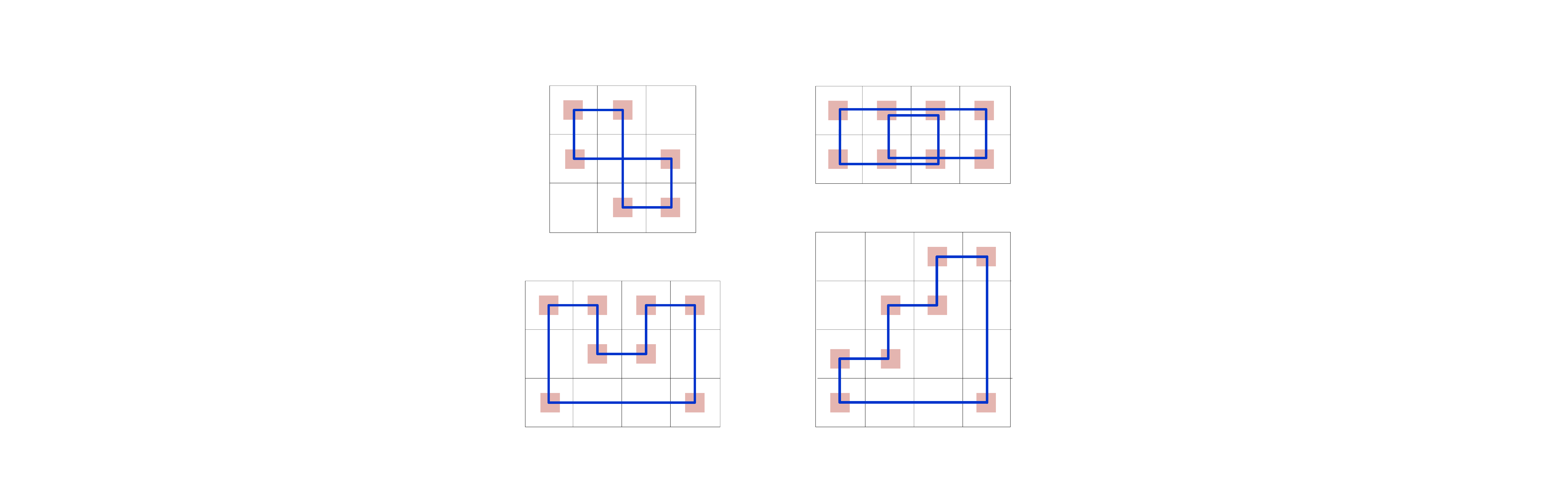}
\caption{The only cycle-$6$ protograph pattern and its candidate (top left) as well as three (out of five) cycle-$8$ protograph patterns and one candidate for each. Light red squares are the non-zero entries of the cycle-candidate.}
\label{cycle_patterns}
\vspace{-1.0em}
\end{figure}

Observe that our optimization variable is $\mathbf{q}$, and the polynomial that involves it is $g(\cdot)$. Therefore, it is both customary and insightful to expand the expectation in \eqref{eqn_thm1} such that the terms with $g(\cdot)$ are separated. We do that for various cycles of interest, cycles-$4$, cycles-$6$, and cycles-$8$, in Lemma~\ref{lemma: obj_fn}. For clarification, the only cycle-$6$ candidate as well as three cycle-$8$ candidates are shown in Fig.~\ref{cycle_patterns} (see also \cite{channel_aware} and \cite{ahh_grade}).

\begin{lemma} \label{lemma: obj_fn}
The expected number of cycle-$4$, cycle-$6$, and cycle-$8$ candidates with distinct entries in the protograph of the previously characterized RMC-SC code (the objective functions of our optimization problems) are given in Equations~\eqref{eq:exp4}, \eqref{eq:exp6}, and \eqref{eq:exp8}, respectively.
\begin{figure*}[ht!]
\noindent\makebox[\linewidth]{\rule{\textwidth}{1.0pt}}
\begin{align}\label{eq:exp4}
&\mathbb{E}[\textup{cycle-$4$}] = \binom{\gamma}{2} \binom{\kappa}{2} 
  \Bigg( r_{\textup{f}}^4 \big[f^2(X) f^2(X^{-1}) \big]_{0}
+\sum_{j=0}^{1} \hspace{+0.3em} \sum_{i=-2m_\textup{f}}^{2m_\textup{f}} \hspace{-0.3em} a_{j}r_{\textup{f}}^{3-j} r_{\textup{n}}^{1+j}  \big[f^2(X) f^{1-j}(X^{-1}) \big]_{i}  \big[g^{1+j}(X^{-1}) \big]_{-i}
\nonumber \\
&+\sum_{j=0}^{1} \hspace{+0.3em} \sum_{i=-m_\textup{f}}^{m_\textup{f}} \hspace{-0.3em} b_{j}r_{\textup{f}}^{2-j} r_{\textup{n}}^{2+j}  \big[f(X) f^{1-j}(X^{-1}) \big]_{i}  \big[g(X) g^{1+j}(X^{-1}) \big]_{-i} 
+ r_{\textup{n}}^4 \big[g^2(X) g^2(X^{-1}) \big]_{0}  \Bigg),
\end{align} 
\textit{where $a_0= 4$, $a_1=2$, $b_0=4$, and $b_1=4$.}
\begin{align}\label{eq:exp6}
&\mathbb{E}[\textup{cycle-$6$}] = 6\binom{\gamma}{3} \binom{\kappa}{3} 
  \Bigg( r_{\textup{f}}^6 \big[f^3(X) f^3(X^{-1}) \big]_{0}
+\sum_{j=0}^{2} \hspace{+0.3em} \sum_{i=-3m_\textup{f}}^{3m_\textup{f}} \hspace{-0.3em}a_{j}r_{\textup{f}}^{5-j} r_{\textup{n}}^{1+j}  \big[f^3(X) f^{2-j}(X^{-1}) \big]_{i}  \big[g^{1+j}(X^{-1}) \big]_{-i}
\nonumber \\
&+\sum_{j=0}^{2} \hspace{+0.3em} \sum_{i=-2m_\textup{f}}^{2m_\textup{f}}\hspace{-0.3em} b_{j}r_{\textup{f}}^{4-j} r_{\textup{n}}^{2+j}  \big[f^2(X) f^{2-j}(X^{-1}) \big]_{i}  \big[g(X) g^{1+j}(X^{-1}) \big]_{-i} \nonumber \\
&+\sum_{j=0}^{1} \hspace{+0.3em} \sum_{i=-m_\textup{f}}^{m_\textup{f}}\hspace{-0.3em} c_{j}r_{\textup{f}}^{2-j} r_{\textup{n}}^{4+j}  \big[f(X) f^{1-j}(X^{-1}) \big]_{i}  \big[g^2(X) g^{2+j}(X^{-1}) \big]_{-i}
+ r_{\textup{n}}^6 \big[g^3(X) g^3(X^{-1}) \big]_{0}  \Bigg),
\end{align} 
\textit{where $a_0= 6$, $a_1=6$, $a_2=2$, $b_0=9$, $b_1=18$, $b_2=6$, $c_0=9$, and $c_1=6$.}
\begin{align}\label{eq:exp8}
&\mathbb{E}[\textup{cycle-$8$}] = A_8
  \Bigg( r_{\textup{f}}^8 \big[f^4(X) f^4(X^{-1}) \big]_{0}
+\sum_{j=0}^{3} \hspace{+0.3em} \sum_{i=-4m_\textup{f}}^{4m_\textup{f}}\hspace{-0.3em} a_{j}r_{\textup{f}}^{7-j} r_{\textup{n}}^{1+j}  \big[f^4(X) f^{3-j}(X^{-1}) \big]_{i}  \big[g^{1+j}(X^{-1}) \big]_{-i}
\nonumber \\
&+\sum_{j=0}^{3} \hspace{+0.3em} \sum_{i=-3m_\textup{f}}^{3m_\textup{f}} \hspace{-0.3em}b_{j}r_{\textup{f}}^{6-j} r_{\textup{n}}^{2+j}  \big[f^3(X) f^{3-j}(X^{-1}) \big]_{i}  \big[g(X) g^{1+j}(X^{-1}) \big]_{-i} \nonumber \\
&+\sum_{j=0}^{2} \hspace{+0.3em} \sum_{i=-2m_\textup{f}}^{2m_\textup{f}} \hspace{-0.3em}c_{j}r_{\textup{f}}^{4-j} r_{\textup{n}}^{4+j}  \big[f^2(X) f^{2-j}(X^{-1}) \big]_{i}  \big[g^2(X) g^{2+j}(X^{-1}) \big]_{-i} \nonumber \\
&+\sum_{j=0}^{1} \hspace{+0.3em} \sum_{i=-m_\textup{f}}^{m_\textup{f}} \hspace{-0.3em}d_{j}r_{\textup{f}}^{2-j} r_{\textup{n}}^{6+j}  \big[f(X) f^{1-j}(X^{-1}) \big]_{i}  \big[g^3(X) g^{3+j}(X^{-1}) \big]_{-i}
+ r_{\textup{n}}^8 \big[g^4(X) g^4(X^{-1}) \big]_{0}  \Bigg),
\end{align} 
\vspace{-0.7em}
\begin{equation}
A_8 = 6\binom{\gamma}{2} \binom{\kappa}{4} 
    + 6\binom{\gamma}{4} \binom{\kappa}{2} 
    + 36\binom{\gamma}{3} \binom{\kappa}{4} 
    + 36\binom{\gamma}{4} \binom{\kappa}{3} 
    + 72\binom{\gamma}{4} \binom{\kappa}{4}, \textit{ where} \nonumber
\end{equation}
\textit{$a_0= 8$, $a_1=12$, $a_2=8$, $a_3=2$, $b_0=16$, $b_1=48$, $b_2=32$, $b_3= 8$, $c_0=36$, $c_1=48$, $c_2=12$, $d_0=16$, and $d_1=8$.}
\noindent\makebox[\linewidth]{\rule{\textwidth}{1.0pt}}
\end{figure*}

\begin{proof}
For cycles-$4$, it can be seen from  Theorem~\ref{thm:elegant objective function} that the expected number of cycle-$4$ candidates is the expansion of the below equation, which is \eqref{eqn_thm1} with $\ell=2$:
\begin{align}
&\mathbb{E}[\textup{cycle-$4$}] = \binom{\gamma}{2} \binom{\kappa}{2} \nonumber \\
&\cdot \Big[ \big( r_{\textup{f}}f(X) +r_{\textup{n}}g(X) \big)^{2}  \big(r_{\textup{f}}f(X^{-1}) +r_{\textup{n}}g(X^{-1}) \big)^{2} \Big]_{0}. 
\end{align}
Alternatively, \eqref{eq:exp4} can also be reached by examining the cases where entries of the cycle-$4$ candidate are distributed among two sets of component matrices, one of which is fixed ($f(X)$ is associated with) and the other is newly added ($g(X)$ is associated with). The entries are also divided into two groups, having $\ell=2$ entries each, with values in $\mathbf{K}$ that have opposite signs in \eqref{eqn_cond}. We define the notation $(\eta_1\textup{--}\eta_2)(\eta_3\textup{--}\eta_4)$ to refer to the situation where $\eta_1$ and $\eta_2$ entries belong to the fixed set with $+$ and $-$ signs, respectively, while $\eta_3$ and $\eta_4$ belong to the new set with $+$ and $-$ signs, respectively. Observe that for the condition in \eqref{eqn_cond} to hold, $\eta_1+\eta_3=\eta_2+\eta_4=2$ (or $\ell$ for a cycle-$2\ell$) must be satisfied. Following this notation, take $(2\textup{--}1)(0\textup{--}1)$ for instance. Three entries will go to the fixed set with probability $r_{\textup{f}}^3$, and one entry will go to the new set with probability $r_{\textup{n}}$. There are $\binom{2}{2} \binom{2}{1} =2$ ways of obtaining this situation. Additionally, $(1\textup{--}2)(1\textup{--}0)$ will be symmetric to this case. Therefore, $a_0$ in \eqref{eq:exp4} will be $4$. The other five situations can be addressed similarly, and they are skipped for brevity.

For cycles-$6$, the same logic can be applied. The expected number of cycle-$6$ candidates is the expansion of the below equation, which is \eqref{eqn_thm1} with $\ell=3$:
\begin{align}
&\mathbb{E}[\textup{cycle-$6$}] = 6\binom{\gamma}{3} \binom{\kappa}{3} \nonumber \\
&\cdot \Big[ \big( r_{\textup{f}}f(X) +r_{\textup{n}}g(X) \big)^{3}  \big(r_{\textup{f}}f(X^{-1}) +r_{\textup{n}}g(X^{-1}) \big)^{3} \Big]_{0}. 
\end{align}
We can use the same alternative method to reach \eqref{eq:exp6}. This time, $\eta_1+\eta_3=\eta_2+\eta_4=3$ must be satisfied. Take $(2\textup{--}2)(1\textup{--}1)$ for instance. Four entries will go to the fixed set with probability $r_{\textup{f}}^4$, and two entries will go to the new set with probability $r_{\textup{n}}^2$. There are $\binom{3}{2} \binom{3}{2} =9$ ways of obtaining this situation. Therefore, $b_0$ in \eqref{eq:exp6} will be $9$. The other nine situations can be addressed similarly.

For cycles-$8$, the same logic can again be applied. The expected number of cycle-$8$ candidates is the expansion of the below equation, which is \eqref{eqn_thm1} with $\ell=4$:
\begin{align}
&\mathbb{E}[\textup{cycle-$8$}] = \nonumber \\
& A_8\Big[ \big( r_{\textup{f}}f(X) +r_{\textup{n}}g(X) \big)^{4}  \big(r_{\textup{f}}f(X^{-1}) +r_{\textup{n}}g(X^{-1}) \big)^{4} \Big]_{0}. 
\end{align}
We can again use the same alternative method to reach \eqref{eq:exp8}. For this case, $\eta_1+\eta_3=\eta_2+\eta_4=4$ must be satisfied. Take $(3\textup{--}2)(1\textup{--}2)$ for instance. Five entries will go to the fixed set with probability $r_{\textup{f}}^5$, and three entries will go to the new set with probability $r_{\textup{n}}^3$. There are $\binom{4}{3}\binom{4}{2} =24$ ways of obtaining this situation. Additionally, $(2\textup{--}3)(2\textup{--}1)$ will be symmetric to this case. Therefore, $b_1$ in \eqref{eq:exp8} will be $48$. The other fourteen situations can be addressed similarly.

Observe that the power of $f(\cdot)$ determines the maximum absolute value $i$ in the relevant summation can take. The combinatorial terms multiplied by $[\cdot]_0$ are the values of $A_{2\ell}$, $\ell$ in $\{2,3,4\}$ (see also \cite{channel_aware} and \cite{ahh_grade}). Finally, note that the total number of $(\eta_1\textup{--}\eta_2)(\eta_3\textup{--}\eta_4)$ unique arrangements for a cycle-$2\ell$ is $\binom{\ell+2}{2}$, which is consistent with (\ref{eq:exp4}), (\ref{eq:exp6}), and (\ref{eq:exp8}).
\end{proof}

\end{lemma}

We highlight that our probabilistic approach for designing RC-SC codes can be extended to the case where redundancy increments are achieved via adding rows instead of component matrices. Consider a $\gamma_{\textup{f}}\times \kappa$ fixed all-one matrix $\mathbf{H}_{\textup{f}}$ and a $\gamma_{\textup{n}}\times \kappa$ new all-one matrix $\mathbf{H}_{\textup{n}}$. The protograph base matrix of the code is the vertical concatenation of $\mathbf{H}_{\textup{f}}$ and $\mathbf{H}_{\textup{n}}$. Now, $\mathbf{p}$ and $\mathbf{q}$ are both of size $1 \times (m+1)$. We redefine
\vspace{-0.1em}\begin{equation}
f(X) = \sum_{j=0}^{m} p_j X^j \textup{ and } g(X) = \hspace{-0.5em}\sum_{j=0}^{m} q_j X^j.
\end{equation}

\begin{lemma} \label{lemma: add_row}
Consider an RC-SC code with memory $m$ where rate compatibility is achieved via adding $\gamma_{\textup{n}} \geq 3$ rows as $\mathbf{H}_{\textup{n}}$ to a fixed protograph base matrix $\mathbf{H}_{\textup{f}}$, $\gamma_{\textup{f}} \geq 3$, as illustrated above. The expected number of cycle-$6$ candidates in the RC-SC protograph (after partitioning, the objective function) is:
\begin{align}
&\mathbb{E}[\textup{cycle-$6$}] 
=6\binom{\gamma_{\textup{f}}}{3}\binom{\kappa}{3}\big[f^3(X) f^3(X^{-1}) \big]_{0} \nonumber \\
&+6\binom{\gamma_{\textup{f}}}{2}\binom{\gamma_{\textup{n}}}{1}\binom{\kappa}{3} \hspace{-0.3em} \sum_{i=-m}^{m} \hspace{-0.4em} \big[f^2(X) f^{2}(X^{-1}) \big]_{i}  \big[g(X) g(X^{-1}) \big]_{-i} \nonumber \\
&+6\binom{\gamma_{\textup{f}}}{1}\binom{\gamma_{\textup{n}}}{2}\binom{\kappa}{3} \hspace{-0.3em} \sum_{i=-m}^{m} \hspace{-0.4em} \big[f(X) f(X^{-1}) \big]_{i}  \big[g^2(X) g^2(X^{-1}) \big]_{-i} \nonumber \\
&+6\binom{\gamma_{\textup{n}}}{3}\binom{\kappa}{3}\big[g^3(X) g^3(X^{-1}) \big]_{0}.
\end{align}

\begin{proof}
In this design, the memory is constant, and entries of a cycle-$6$ candidate are distributed among the fixed rows and the newly added rows. We redefine the notation $(\eta_1\textup{--}\eta_2)(\eta_3\textup{--}\eta_4)$ to refer to the situation where $\eta_1$ and $\eta_2$ entries belong to the fixed rows with $+$ and $-$ signs, respectively, while $\eta_3$ and $\eta_4$ belong to the new rows with $+$ and $-$ signs, respectively. For the condition in \eqref{eqn_cond} to hold, $\eta_1+\eta_3=\eta_2+\eta_4=3$ must be satisfied. On the cycle candidate, each entry with a $+$ sign must have an entry with a $-$ sign in the same row, which means they follow the same distribution (either $\mathbf{p}$ or $\mathbf{q}$). Thus, $\eta_1=\eta_2$ and $\eta_3=\eta_4$ must also be satisfied here. There are $\eta_1$ rows selected from the fixed rows and $\eta_3$ rows selected from the new rows to have all $6$ entries of the cycle candidate. Thus, we have the combinatorial term $\binom{\gamma_{\textup{f}}}{\eta_1}\binom{\gamma_{\textup{n}}}{\eta_3}$, in four situations, multiplied by $\binom{\kappa}{3}$ to select $3$ columns. There are $6$ ways to order these $3$ columns. Observe that for a cycle-$6$, the maximum absolute value $i$ in the summations can take for a non-zero term is $m$ here. The multiplied polynomial coefficients $[\cdot]_i [\cdot]_{-i}$ are to find the constant polynomial terms, which are the ones of interest to compute the expectation.
\end{proof}

\end{lemma}

\vspace{-0.7em}
\begin{remark}
Rate compatibility with constant memory can also be extended to the case of adding columns (rate increments). This approach applies to cycles-$4$ and cycles-$8$ as well.
\end{remark}

In the remainder of the paper, we return to focus on RMC-SC codes, and the notation, for example $f(X)$ and $g(X)$, is reverted to the one used for RMC-SC codes.

\section{Gradient-Descent Distribution}\label{sec_grad}

In this section, we introduce our optimization problem, its solution form, and the gradients. Then, we propose a gradient-descent algorithm that finds the distribution $\mathbf{q}$.

We focus on specific cycles-$2\ell$ in $\mathbf{H}$. The extension to a weighted combination of cycle-count expectations is simple. For the RMC-SC code protograph, our optimization problem, where the domain is $\mathbf{q}$ in $[0,1]^{m_{\textup{n}}}$, is formulated as follows:
\begin{equation}
\underset{\mathbf{q}}{\textup{minimize}} \textup{ } \textup{ } \mathbb{E}[\textup{cycle-}2\ell] = E_{2\ell}(\mathbf{q}) \nonumber
\end{equation}
\vspace{-1.5em}
\begin{equation}
\textup{subject to } \textup{ } \sum_{j=m_{\textup{f}}+1}^{m_{\textup{f}}+m_{\textup{n}}}q_j=1.
\end{equation}
Note that ``fixed'' in this section still means distribution-wise fixed. The notation $\mathbf{1}_u$ means an all-one vector of length $u$.

\begin{theorem} \label{thm:solution}
Consider an RMC-SC code that has $\mathbf{H}_{\textup{f}}$, with $r_{\textup{f}}$ and $m_{\textup{f}}$, as well as $\mathbf{H}_{\textup{n}}$, with $r_{\textup{n}}$ and $m_{\textup{n}}$. Then, for $\mathbb{E}[\textup{cycle-}2\ell]$ to reach its local minimum, the following equation holds for some $c \in \mathbb{R}_{>0}$:
\begin{align}
& A_{2\ell} \Big[ \big(r_{\textup{f}}f(X) + r_{\textup{n}}g(X)\big)^{\ell} 
      \big(r_{\textup{f}}f(X^{-1}) + r_{\textup{n}}g(X^{-1})\big)^{\ell-1} \Big]_{i}  \nonumber \\
&= \frac{c}{2\ell r_{\textup{n}}}, \ \forall i \in \{m_{\textup{f}}+1,m_{\textup{f}}+2,\dots,m_{\textup{f}}+m_{\textup{n}}\}. \label{sol_form}
\end{align}

\begin{proof}
Our optimization problem satisfies the linear independence constraint qualification (LICQ), which implies that we can apply Karush-Kuhn-Tucker (KKT) conditions. Hence, we consider the Lagrangian
\vspace{-0.7em}
\begin{equation}
L_{2\ell}(\mathbf{q})=\mathbb{E}[\textup{cycle-}2\ell]+c\Big(1-\sum_{j=m_{\textup{f}}+1}^{m_{\textup{f}}+m_{\textup{n}}}q_j\Big).
\end{equation}
Using \eqref{eqn_thm1}, the gradient of $L_{2\ell}(\mathbf{q})$ is computed as follows:
\vspace{-0.5em}
\begin{align} \label{gradient_2ell}
&\nabla_{\mathbf{q}} L_{2\ell}(\mathbf{q})
=\nabla_{\mathbf{q}}\Big(\mathbb{E}[\textup{cycle-}2\ell]+c\Big(1- \sum_{j=m_{\textup{f}}+1}^{m_{\textup{f}}+m_{\textup{n}}}q_j\Big)\Big) \nonumber \\
&\hspace{-0.7em}=\nabla_{\mathbf{q}}\Big( A_{2\ell}\Big[ \big( r_{\textup{f}}f(X) +r_{\textup{n}}g(X) \big)^{\ell}  \big(r_{\textup{f}}f(X^{-1}) +r_{\textup{n}}g(X^{-1}) \big)^{\ell} \Big]_{0}         \nonumber \\
&\hspace{2.4em}+ c\Big(1- \sum_{j=m_{\textup{f}}+1}^{m_{\textup{f}}+m_{\textup{n}}}q_j\Big)\Big)         \nonumber \\
&\hspace{-0.7em}=A_{2\ell}\Big[ \big( r_{\textup{f}}f(X) +r_{\textup{n}}g(X) \big)^{\ell-1}  \big(r_{\textup{f}}f(X^{-1}) +r_{\textup{n}}g(X^{-1}) \big)^{\ell }    \nonumber\\
&\hspace{3.1em} \cdot \ell r_{\textup{n}}\nabla_{\mathbf{q}} g(X)   \Big]_{0} \nonumber \\
&\hspace{-0.7em}+A_{2\ell}\Big[ \big( r_{\textup{f}}f(X) +r_{\textup{n}}g(X) \big)^{\ell}  \big(r_{\textup{f}}f(X^{-1}) +r_{\textup{n}}g(X^{-1}) \big)^{\ell -1}   \nonumber \\
&\hspace{3.1em} \cdot \ell r_{\textup{n}}\nabla_{\mathbf{q}} g(X^{-1})   \Big]_{0} -c\mathbf{1}_{m_{\textup{n}}} \nonumber \\
&\hspace{-0.7em}=A_{2\ell}2\ell r_{\textup{n}} \Big[ \big( r_{\textup{f}}f(X) +r_{\textup{n}}g(X) \big)^{\ell}  \big(r_{\textup{f}}f(X^{-1}) +r_{\textup{n}}g(X^{-1}) \big)^{\ell -1} \nonumber \\
& \hspace{0.5em} \cdot \big[ X^{-(m_{\textup{f}} +1)}\ X^{-(m_{\textup{f}} +2)}\ \dots \ X^{-(m_{\textup{f}}+m_{\textup{n}})}\big]\Big]_{0} -c\mathbf{1}_{m_{\textup{n}}}.
\end{align}
$\mathbb{E}[\textup{cycle-}2\ell]$  reaches its local minimum when $\nabla_{\mathbf{q}}L_{2\ell}(\mathbf{q}) =\mathbf{0}_{m_{\textup{n}}}$, which results in \eqref{sol_form}.
\end{proof}
 
\end{theorem}
\vspace{-0.2em}

Once again, it is both customary and insightful to examine the detailed solution form when the terms with $g(\cdot)$ are separated. We do that for cycles-$4$ in Lemma~\ref{gradient_cycle_4}.

\begin{lemma} \label{gradient_cycle_4}
For $\mathbb{E}[\textup{cycle-$4$}]$ associated with the protograph of the previously characterized RMC-SC code to reach its local minimum, Equation~\eqref{sol_form_4} holds for some $c\in \mathbb{R}_{>0}$.
\begin{figure*}[ht!]
\noindent\makebox[\linewidth]{\rule{\textwidth}{1.0pt}}
\begin{align}
&\binom{\gamma}{2} \binom{\kappa}{2} \Bigg( 
\sum_{j=0}^{1} \hspace{+0.3em} \sum_{i=-2m_\textup{f}}^{2m_\textup{f}} 
\hspace{-0.3em} a_{j} r_{\textup{f}}^{3-j} r_{\textup{n}}^{1+j}   
\big[f^2(X) f^{1-j}(X^{-1}) \big]_{i} 
\begin{bmatrix} 
[(1+j)g^{j}(X^{-1})]_{-i+m_{\textup{f}} +1}\\ 
[(1+j)g^{j}(X^{-1})]_{-i+m_{\textup{f}} +2}\\ 
\vdots \\ 
[(1+j)g^{j}(X^{-1})]_{-i+m_{\textup{f}}+m_{\textup{n}}}\\  
\end{bmatrix}^\mathrm{T}  
\nonumber \\
&+\sum_{j=0}^{1} \hspace{+0.3em} \sum_{i=-m_\textup{f}}^{m_\textup{f}} 
\hspace{-0.3em} b_{j} r_{\textup{f}}^{2-j} r_{\textup{n}}^{2+j}  
\big[f(X) f^{1-j}(X^{-1}) \big]_{i} 
\begin{bmatrix} 
[g^{1+j}(X^{-1})]_{-i-m_{\textup{f}} -1} 
+ [(1+j)g(X)g^{j}(X^{-1})]_{-i+m_{\textup{f}}+1}\\ 
[g^{1+j}(X^{-1})]_{-i-m_{\textup{f}} -2} 
+ [(1+j)g(X)g^{j}(X^{-1})]_{-i+m_{\textup{f}}+2}\\ 
\vdots \\ 
[g^{1+j}(X^{-1})]_{-i-m_{\textup{f}}-m_{\textup{n}}} 
+ [(1+j)g(X)g^{j}(X^{-1})]_{-i+m_{\textup{f}}+m_{\textup{n}}}
\end{bmatrix}^\mathrm{T} \nonumber \\
&+r_{\textup{n}}^{4} \cdot 
\begin{bmatrix} 
[4g^{2}(X)g(X^{-1})]_{m_{\textup{f}} +1} \\ 
[4g^{2}(X)g(X^{-1})]_{m_{\textup{f}} +2}\\ 
\vdots \\ 
[4g^{2}(X)g(X^{-1})]_{m_{\textup{f}}+m_{\textup{n}}}  
\end{bmatrix}^\mathrm{T} \Bigg )
= c\mathbf{1}_{m_{\textup{n}}},  \label{sol_form_4}
\end{align}
\textit{where $a_0= 4$, $a_1=2$, $b_0=4$, and $b_1=4$.} \\
\noindent\makebox[\linewidth]{\rule{\textwidth}{1.0pt}}
\end{figure*}
\begin{proof}
Consider the Lagrangian $L_{4}(\mathbf{q})=\mathbb{E}[\textup{cycle-$4$}]+c\big(1- \sum_{j=m_{\textup{f}}+1}^{m_{\textup{f}}+m_{\textup{n}}}q_j\big)$. Using \eqref{eq:exp4}, the gradient of $L_{4}(\mathbf{q})$ is computed as demonstrated in \eqref{eq_gradproof4}.
\begin{figure*}[ht!]
\vspace{-0.3em}
\noindent\makebox[\linewidth]{\rule{\textwidth}{1.0pt}}
\begin{align}\label{eq_gradproof4}
&\nabla_{\mathbf{q}} L_{4}(\mathbf{q})
=\nabla_{\mathbf{q}} L_{4}\Big(\mathbb{E}[\textup{cycle-$4$}] + c\Big(1-\sum_{j=m_{\textup{f}}+1}^{m_{\textup{f}}+m_{\textup{n}}}q_j\Big)\Big)
=\nabla_{\mathbf{q}} L_{4}\big(\mathbb{E}[\textup{cycle-$4$}]\big)-c\mathbf{1}_{m_{\textup{n}}} \nonumber \\
&=\nabla_{\mathbf{q}}\Bigg( \binom{\gamma}{2} \binom{\kappa}{2} 
  \Big( r_{\textup{f}}^4 \big[f^2(X) f^2(X^{-1}) \big]_{0}
+\sum_{j=0}^{1} \hspace{+0.3em} \sum_{i=-2m_\textup{f}}^{2m_\textup{f}} \hspace{-0.3em} a_{j}r_{\textup{f}}^{3-j} r_{\textup{n}}^{1+j}  \big[f^2(X) f^{1-j}(X^{-1}) \big]_{i}  \big[g^{1+j}(X^{-1}) \big]_{-i}
\nonumber \\
&+\sum_{j=0}^{1} \hspace{+0.3em} \sum_{i=-m_\textup{f}}^{m_\textup{f}} \hspace{-0.3em} b_{j}r_{\textup{f}}^{2-j} r_{\textup{n}}^{2+j}  \big[f(X) f^{1-j}(X^{-1}) \big]_{i}  \big[g(X) g^{1+j}(X^{-1}) \big]_{-i}
+ r_{\textup{n}}^4 \big[g^2(X) g^2(X^{-1}) \big]_{0}  \Bigg) -c\mathbf{1}_{m_{\textup{n}}} \nonumber \\
&=\binom{\gamma}{2} \binom{\kappa}{2} \Bigg( 
\sum_{j=0}^{1} \hspace{+0.3em} \sum_{i=-2m_\textup{f}}^{2m_\textup{f}} 
\hspace{-0.3em} a_{j} r_{\textup{f}}^{3-j} r_{\textup{n}}^{1+j}   
\big[f^2(X) f^{1-j}(X^{-1}) \big]_{i}
\Big[(1+j)g^{j}(X^{-1})\big[ X^{-(m_{\textup{f}} +1)}\ X^{-(m_{\textup{f}} +2)}\ \dots \ X^{-(m_{\textup{f}}+m_{\textup{n}})}\big]\Big]_{-i} \nonumber \\
&+\sum_{j=0}^{1} \hspace{+0.3em} \sum_{i=-m_\textup{f}}^{m_\textup{f}} 
\hspace{-0.3em} b_{j} r_{\textup{f}}^{2-j} r_{\textup{n}}^{2+j} \big[f(X) f^{1-j}(X^{-1}) \big]_{i} \nonumber \\
& \hspace{3.0em}\cdot\Big[g^{1+j}(X^{-1})\big[ X^{m_{\textup{f}} +1}\ X^{m_{\textup{f}} +2}\ \dots \ X^{m_{\textup{f}}+m_{\textup{n}}}\big]
+ (1+j)g(X)g^{j}(X^{-1})\big[ X^{-(m_{\textup{f}} +1)}\ X^{-(m_{\textup{f}} +2)}\ \dots \ X^{-(m_{\textup{f}}+m_{\textup{n}})}\big] \Big]_{-i} \nonumber \\
&+r_{\textup{n}}^{4} 
\Big[4g^{2}(X)g(X^{-1})\big[ X^{-(m_{\textup{f}} +1)}\ X^{-(m_{\textup{f}} +2)}\ \dots \ X^{-(m_{\textup{f}}+m_{\textup{n}})}\big]\Big]_0  
 \Bigg )
- c\mathbf{1}_{m_{\textup{n}}}.
\end{align}
\noindent\makebox[\linewidth]{\rule{\textwidth}{1.0pt}}
\end{figure*}
$\mathbb{E}[\textup{cycle-$4$}]$ reaches its local minimum when $\nabla_{\mathbf{q}}L_{4}(\mathbf{q}) =\mathbf{0}_{m_{\textup{n}}}$, which results in \eqref{sol_form_4}.
\end{proof}

\end{lemma}

\begin{remark}
This approach with detailed gradients can also be extended to cycles-$6$ and cycles-$8$ in a similar manner. We skip the details for simplicity.
\end{remark}

\begin{algorithm}[]
\caption{Rate-Memory-Compatible Gradient-Descent Distributor (RMC-GRADE) for Cycle Optimization} \label{algo: RMC_GD}
\begin{algorithmic}[1]
\Statex \textbf{Inputs:} $\gamma,\kappa, m_{\textup{f}}, m_{\textup{n}}, r_{\textup{f}}, r_{\textup{n}}$: parameters of the RMC-SC code; $\mathbf{p}$: fixed distribution obtained in the previous design stage; $w_6$: cycle-$6$ weight; $w_8$: cycle-$8$ weight; $\epsilon, \alpha$: accuracy and step size of gradient descent.
\Statex \textbf{Outputs:} $\mathbf{q}$: locally-optimal probability-distribution vector for the RMC-SC code; $E_6$: the expected number of cycle-$6$ candidates in the RMC-SC protograph; $E_8$: the expected number of cycle-$8$ candidates in the RMC-SC protograph.
\Statex \textbf{Intermediate Variables:} $F_{\textup{prev}}, F_{\textup{cur}}$: values of the objective function at the previous and current iterations; $\mathbf{g}$: the gradient vector (of size $1 \times m_{\textup{n}}$) of the objective function.
\State $F_{\textup{prev}}=1$, $F_{\textup{cur}}=1$.
\State $F_{\textup{prev}}=F_{\textup{cur}}$.
\State Compute $E_6=\mathbb{E}[\textup{cycle-$6$}]$ as given in (\ref{eq:exp6}).
\State Compute $E_8=\mathbb{E}[\textup{cycle-$8$}]$ as given in (\ref{eq:exp8}).
\State $F_{\textup{cur}}=w_6E_6+w_8E_8$.
\State Compute the gradient of $F_{\textup{cur}}$ as $\mathbf{g}=\nabla_{\mathbf{q}}F_{\textup{cur}}=w_6\nabla_{\mathbf{q}}E_6+w_8\nabla_{\mathbf{q}}E_8$ using Theorem~\ref{thm:solution}, $\mathbf{g} \gets \mathbf{g} - \mathrm{mean}(\mathbf{g})\mathbf{1}_{m_{\textup{n}}}$.
\If{$(|F_{\textup{cur}}-F_{\textup{prev}}| > \epsilon)$} 
\State $\mathbf{q} \gets \mathbf{q} - \alpha \displaystyle \frac{\mathbf{g}}{\| \mathbf{g}\|}$.
\State \textbf{go to} Step 2.
\EndIf
\State \textbf{return} $\mathbf{q}$, $E_6$, and $E_8$.
\end{algorithmic}
\end{algorithm}

Next, we develop a gradient-descent algorithm that obtains a locally-optimal probability distribution for the new components, $\mathbf{q}$, of an RMC-SC code to minimize the expected number of short cycles, cycles-$6$ and cycles-$8$, or a combination of such expectations using the concepts in \cite{ahh_grade}.

The expected number of cycle-$6$ candidates in the protograph is $\mathbb{E}[\textup{cycle-$6$}]$, which is given in (\ref{eq:exp6}). The expected number of cycle-$8$ candidates in the protograph is $\mathbb{E}[\textup{cycle-$8$}]$, which is given in (\ref{eq:exp8}). Our objective is to minimize the weighted sum $w_6\mathbb{E}[\textup{cycle-$6$}]+w_8\mathbb{E}[\textup{cycle-$8$}]$, where $w_6$ is the weight of cycles-$6$ and $w_8$ is the weight of cycles-$8$. One can choose $w_8$ from $\{0,1\}$, depending on the expected population of cycles-$6$. Since shorter cycles tend to be more detrimental than longer ones, it is logical to set $w_6>1$ when $w_8 = 1$. While the weights can be dynamically adjusted for different design stages (see Section~\ref{sec_mcmc}), we use the same weights for all design stages in our gradient descent algorithm.

Initially, we obtain the locally-optimal probability distribution $\mathbf{p}^*$ of length $m_s+1$ for an SC code with $m_s=\sum_{d=0}^s m_{\textup{n},d}$ using the method given in \cite{ahh_grade}. Moreover, we compute the probabilities $r_{\textup{n},0}=\sum_{j=0}^{m_{\textup{n},0}} \mathbf{p}^*_j$ and $r_{\textup{n},d}=\sum_{j=m_{\textup{f},d}+1}^{m_{\textup{f},d}+m_{\textup{n},d}} \mathbf{p}^*_j$, for all $d$ in $\{1,2,\dots,s\}$. Design Stage~0 is simple as illustrated in Section~\ref{sec_design}. Consider now a generic design stage indexed by $d$, for all $d$ in $\{1,2,\dots,s\}$, and again drop $d$ from the subscripts. To obtain the locally-optimal probability distribution $\mathbf{q}$ and then $\mathbf{u}$ for the respective RMC-SC code, we apply these steps:
\begin{itemize}
	\item We use the locally-optimal final probability distribution obtained in Design Stage~$d-1$ of length $m_{\textup{f}}+1$, the previous $\mathbf{u}$, as the fixed distribution $\mathbf{p}$ in Design Stage~$d$.
	\item We obtain the probability distribution $\mathbf{q}$ using Algorithm~\ref{algo: RMC_GD}, which we call the RMC-GRADE algorithm. Recall that the final distribution for the RMC-SC code at Design Stage~$d$ is:
\begin{align}
\mathbf{u} = \left[ \frac{r_{\textup{f}}}{r_{\textup{f}}+r_{\textup{n}}}\mathbf{p} \textup{ }\textup{ }\textup{ } \frac{r_{\textup{n}}}{r_{\textup{f}}+r_{\textup{n}}}\mathbf{q} \right] \in [0,1]^{1\times(m_{\textup{f}}+m_{\textup{n}}+1)}, \nonumber
\end{align}
which will be the distribution $\mathbf{p}$ in the next design stage.
\end{itemize}
For Design Stage~0, $\mathbf{u}=\mathbf{q}$ is the locally-optimal probability distribution for an SC code with memory $m_{\textup{n},0}$ (nothing fixed).

\begin{example} \label{example:GD}
Consider an RMC-SC code with parameters $\gamma=7$, $\kappa=35$, $m_{\textup{n},0}=8$, $m_{\textup{n},1}=3$, and $m_{\textup{n},2}=4$. Using Algorithm~\ref{algo: RMC_GD} results in the following:
\begin{itemize}
    \item At Design Stage~0, the final distribution is $\mathbf{u}=[0.2494 \ 0.0925 \ 0.0685 \ 0.0605 \ 0.0582 \ 0.0604 \ 0.0688 \allowbreak 0.0920 \ 0.2497]$. This results in $\mathbb{E}[\textup{cycle-$6$}] \approx 1{,}580$ and $\mathbb{E}[\textup{cycle-$8$}] \approx 35{,}193$ (rounded).
    \item At Design Stage~1, the final distribution is $\mathbf{u}=[0.2086 \ 0.0774 \ 0.0573 \ 0.0506 \ 0.0487 \ 0.0506 \ 0.0576 \allowbreak 0.0769 \ 0.2088 \ 0.0018 \ 0.0472 \ 0.1148]$. This results in $\mathbb{E}[\textup{cycle-$6$}] \approx 3{,}780$ and $\mathbb{E}[\textup{cycle-$8$}] \approx 202{,}422$ (rounded).
    \item At Design Stage~2, the final distribution is $\mathbf{u}=[0.1324 \ 0.0491 \ 0.0364 \ 0.0321 \ 0.0309 \ 0.0321 \ 0.0365 \allowbreak 0.0488 \ 0.1325 \ 0.0011 \ 0.0300 \ 0.0729 \ 0.0004 \ 0.0691 \allowbreak 0.0863 \ 0.2096]$. This results in $\mathbb{E}[\textup{cycle-$6$}] \approx 41{,}201$ and $\mathbb{E}[\textup{cycle-$8$}] \approx 5{,}431{,}700$ (rounded).
\end{itemize}
\end{example}

\begin{figure}
\center
\hspace{-0.5em}\includegraphics[trim={0.0in 0.7in 0.0in 0.8in}, width=3.55in]{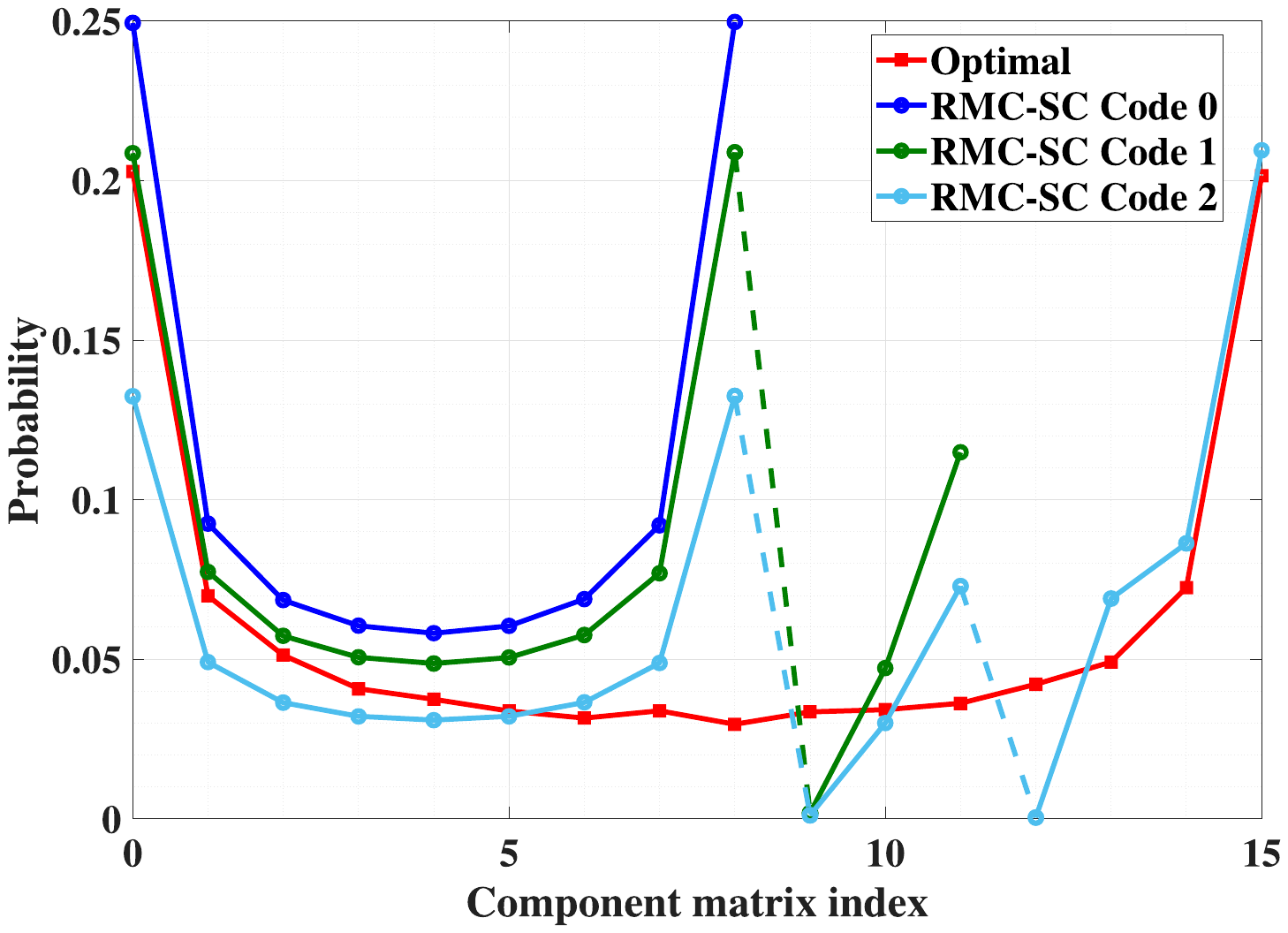}
\caption{Probability distributions of RMC-SC codes at different design stages as well as the optimal distribution for $m=15$.}
\label{fig:gd_dist}
\end{figure}

In Fig.~\ref{fig:gd_dist}, we show the three RMC-SC final distributions at different design stages that are listed in Example~\ref{example:GD} as well as the optimal edge distribution of an SC code with memory $m=15$. For RMC-SC Code~0, this U-shaped distribution, where the probability is higher for component matrices closer to the two edges, is expected to be the optimal according to \cite{ahh_grade}. Every time we add new component matrices, Algorithm~\ref{algo: RMC_GD} naturally attempts to follow the same trend to minimize cycle counts. Therefore, the optimal distribution $\mathbf{q}$ is always such that the probability increases with component matrix index as shown in the figure for RMC-SC Codes~1 and 2. Observe that the depicted diversity in probability distributions of our RMC-SC codes contributes to remarkable performance gains as we shall see in Section~\ref{sec_exp}. Recall that optimality here is always with respect to cycle count.

\section{Finite-Length MCMC Optimization}\label{sec_mcmc}

\begin{algorithm}
\caption{Markov Chain Monte Carlo (MC\textsuperscript{2}) Optimizer for Cycle-Count Reduction in RMC-SC Codes} \label{algo:MC}
\begin{algorithmic}[1]
\Statex \textbf{Inputs:} $\mathcal{L}$: list of cycle candidates of interest; $b$: number of entries to be simultaneously updated; $\mathcal{N}$: set of vectors that do not have any fixed entry altered; $\mathbf{x}_{\text{init}}$: initial input vector; $\mathcal{T}$: maximum transition count (maximum number of iterations); $\mathbf{a}$: set of possible values for each entry of the input vector.
\Statex \textbf{Outputs:} $\overline{C}_{\text{opt}}$: minimum value of the objective function (normalized cycle count) recorded during iterations; $\mathbf{x}_{\text{opt}}$: optimal value of the input vector resulting in the minimum objective function; $i$: number of transitions.
\Statex \textbf{Intermediate Variables:} $S$: list of index $b$-tuples; $\mathbf{x}$: input vector, a concatenation of the rows of the partitioning matrix or the lifting matrix depending on the~MC\textsuperscript{2} phase; $\nu$: index tuple of entries that are currently processed; $Z$: normalizing factor; $\mathbf{x}'$: input vector used for intermediate calculations; $\mathbf{x}'_{\nu}$, $\mathbf{x}_{\setminus \nu}$: vectors of entries in $\mathbf{x}'$ and $\mathbf{x}$ that are indexed by $\nu$ and $\{0,1,\dots,\gamma\kappa-1\} \setminus \nu$, respectively; $P_\nu$, $P_\nu^*$: normalized and non-normalized mappings of conditional probability mass function values associated with the transition between Markov chain states $\mathbf{x}$ and $\mathbf{x}'$, where $\mathbf{x}'$ and $\mathbf{x}$ differ only at entries indexed by $\nu$; $\overline{C}(\mathbf{x}')$: normalized count at $\mathbf{x}'$.

\State Initialize $S$ by going over $\mathcal{L}$ and calculating the common cycle counts for all entry pairs, then determine $b$ indices of the most correlated entries for each entry. 
\State $\mathbf{x} \gets \mathbf{x}_{\text{init}}$, $\overline{C}_{\text{opt}} \gets 1$, $i \gets 0$.
\While{$i < \mathcal{T}$}
	\State Shuffle the order of $S$ randomly.
    \ForEach{$\nu \in S$}
        \State $i \gets i+1$, $Z \gets 0$.
        \ForEach{$\mathbf{x}'_{\nu} \in \mathbf{a}^b$}
            \State Merge $\mathbf{x}'_{\nu}$ and $\mathbf{x}_{\setminus \nu}$ to reach $\mathbf{x}'$.
            \If {$\mathbf{x}'$ $\in \mathcal{N}$}
                \If {$\mathbf{x}'$ does not satisfy the constraints}
                	\State $P_\nu(\mathbf{x}') \gets 0$.
                \Else \textit{ // Norm constraints are satisfied for partitioning and shorter cycle counts remain $0$'s for lifting.}
                    \State Evaluate $\overline{C}(\mathbf{x}')$.
                    \If{$\overline{C}(\mathbf{x}') < \overline{C}_{\text{opt}}$}
                        \State $\overline{C}_{\text{opt}}\gets \overline{C}(\mathbf{x}')$, $\mathbf{x}_{\text{opt}}\gets \mathbf{x}'$.
                        \If{$\overline{C}_{\text{opt}} = 0$} \textbf{go to} Step~28.
                        \EndIf
                    \EndIf
                    \State Update $P_\nu^*(\mathbf{x}')$, $Z\gets Z+P_\nu^*(\mathbf{x}')$. \textit{// Transition probabilities decay exponentially with $\overline{C}(\mathbf{x}')$.}
                \EndIf
            \EndIf
        \EndFor
        \State $P_\nu\gets P_\nu^*/Z$. \textit{// Normalization to reach a probability distribution, for all $\mathbf{x}$.}
        \State Sample from the distribution $P_\nu$ to update $\mathbf{x}$.
    \EndFor
    \State Update MC\textsuperscript{2} internal and distribution variables. \textit{// This is to control the transition rate.}
\EndWhile
\State \textbf{return} $\overline{C}_{\text{opt}}$, $\mathbf{x}_{\text{opt}}$, and $i$.
\end{algorithmic}
\end{algorithm}

We now summarize our MC\textsuperscript{2} algorithm, which is an updated version of that in \cite{reins_mcmc}. This MC\textsuperscript{2} algorithm is a finite-length algorithmic optimizer (FL-AO) that we use to obtain the partitioning and lifting matrices of our RMC-SC codes. The MC\textsuperscript{2} algorithm is used twice here. First, we find a locally-optimal partitioning matrix that minimizes the weighted sum of short cycle candidate counts in the RMC-SC protograph. Second, we find a locally-optimal lifting matrix that minimizes short cycle counts in the final RMC-SC code. The algorithm iterates until cycle counts are not reduced further or until the given iteration limit is reached.

We first focus on the MC\textsuperscript{2} partitioning phase. Before we start designing RMC-SC codes in a recursive manner, we obtain the partitioning matrix $\mathbf{K}^*$ for an SC code with memory $m_s$, where $m_s=\sum_{d=0}^s m_{\textup{n},d}$, using the MC\textsuperscript{2} algorithm. The reason to find $\mathbf{K}^*$ first is to determine the protograph base matrices of our RMC-SC codes. We have a VN-degree condition (restriction) in our RMC-SC codes, which is that for each RMC-SC code, the minimum VN degree is $3$. For a generic design stage indexed by $d$, for all $d$ in $\{1,2,\dots,s\}$, the base matrix $\mathbf{H}_{\textup{n}}$ is obtained as follows. If $\mathbf{K}^*(i,j)>m_{\textup{f}}$ and $\mathbf{K}^*(i,j)<m_{\textup{f}}+m_{\textup{n}}+1$, then $\mathbf{H}_{\textup{n}}(i,j)=1$; otherwise, $\mathbf{H}_{\textup{n}}(i,j)=0$. Next, $\mathbf{H}_{\textup{n}}$ is partitioned into $m_{\textup{n}}$ disjoint components $\mathbf{H}_k, \textup{ } k = m_{\textup{f}}+1, {m_{\textup{f}}+2}, \dots, m_{\textup{f}}+m_{\textup{n}}$, where $\sum_{k=m_{\textup{f}}+1}^{m_{\textup{f}}+m_{\textup{n}}} \mathbf{H}_k = \mathbf{H}_{\textup{n}}$. For Design Stage~0, $\mathbf{H}_{\textup{n}}$ is determined as follows. If $\mathbf{K}^*(i,j)<m_{\textup{n}}+1$, then $\mathbf{H}_{\textup{n}}(i,j)=1$; otherwise, $\mathbf{H}_{\textup{n}}(i,j)=0$. Next, $\mathbf{H}_{\textup{n}}$ is partitioned into $m_{\textup{n}}+1$ disjoint components $\mathbf{H}_k, \textup{ } k = 0, 1, \dots, m_{\textup{n}}$, where $\sum_{k=0}^{m_{\textup{n}}} \mathbf{H}_k = \mathbf{H}_{\textup{n}}$. We dropped $d$ from the subscripts, including $d=0$.

We find the locally-optimal partitioning matrix $\mathbf{K}$ and then lifting matrix $\mathbf{T}$ of each RMC-SC code using Algorithm~\ref{algo:MC}. For a generic design stage indexed by $d$, for all $d$ in $\{1,2,\dots,s\}$, $\mathbf{K}$ and $\mathbf{T}$ have fixed and optimizable entries. The optimizable entries are the entries corresponding to those in $\mathbf{H}_{\textup{n}}$ that are $1$'s. The fixed entries are the entries that are decided by the previous runs of the MC\textsuperscript{2} algorithm in earlier design stages, which are the entries corresponding to those in $\mathbf{H}_{\textup{f}}$ that are $1$'s. For Design Stage~0, all non-negative entries are optimizable entries in both $\mathbf{K}$ and $\mathbf{T}$. For Design Stage~$d$, for all $d$ in $\{0,1,\dots,s-1\}$, there are also entries that will be optimized in upcoming design stages, which we set to $-1$ in both $\mathbf{K}$ and $\mathbf{T}$. While initializing the input state vector of the of the MC\textsuperscript{2} algorithm at the relevant stage, we assign its relevant entries to the fixed values determined in earlier design stages.

In the partitioning phase of the MC\textsuperscript{2} algorithm, we obtain the partitioning matrices of our RMC-SC codes in a recursive manner. The algorithm works on the set of short cycle candidates in the protograph base matrix and aims to find a locally-optimal partitioning matrix $\mathbf{K}$ that minimizes the weighted sum of the number of active candidates after partitioning. We use code-dependent weighting in our MC\textsuperscript{2} partitioning phase, i.e., the weighting mechanism changes according to the design stage. The MC\textsuperscript{2} algorithm begins with an initial random partitioning matrix, which has fixed entries and optimizable entries, whose edge distribution approximately matches that produced by the RMC-GRADE algorithm in order to reduce the MC\textsuperscript{2} search space \cite{reins_mcmc}. Furthermore, we restrict the state vectors in the Markov chain such that both their $L_1$ and $L_{\infty}$ distances from the initial state vector are bounded, ensuring that the algorithm produces outcomes whose distributions remain close to the RMC-GRADE distribution, which is the locally-optimal distribution.

In the lifting phase of the MC\textsuperscript{2} algorithm, we obtain~the lifting matrices of our RMC-SC codes in a recursive manner that is similar to the partitioning phase. Algorithm~\ref{algo:MC} operates on the set of short cycle candidates in the RMC-SC protograph and tries to find a locally-optimal lifting matrix $\mathbf{T}$ that minimizes the number of active cycles after lifting. The MC\textsuperscript{2} algorithm is initialized with a lifting matrix, which has fixed entries and optimizable entries, that results in removing all cycles-$4$. Observe that removing all cycles-$4$ during lifting is typically not a challenging task. Subsequently, the algorithm tries to minimize the number of cycles-$6$ in the code graph, and when they are removed entirely, it addresses cycles-$8$ while ensuring that shorter cycles remain collectively eliminated.

\section{Experimental Results}\label{sec_exp}

In this section, we present numerical results and performance plots of the proposed RMC-SC codes. We also show the significant gains achieved by our RMC-SC codes over codes designed via a straightforward approach that optimizes one SC code only then consecutively reduces its memory.

Straightforward-SC (SF-SC) codes are designed as follows. First, a locally-optimal probability distribution for the SC code with memory $m_s$ is determined \cite{ahh_grade}. Using this distribution, the partitioning matrix $\mathbf{K}^*$ is derived using the MC\textsuperscript{2} algorithm. Subsequently, the lifting matrix $\mathbf{T}^*$ for this SC code is also derived using the MC\textsuperscript{2} algorithm. The resulting code is the SF-SC code with memory $m_{\textup{f},s}+m_{\textup{n},s}=m_s$. To obtain the SF-SC code with memory $m_{\textup{f},d}+m_{\textup{n},d}$, for all~$d$ in $\{1,2,\dots,s-1\}$, we obtain the relevant partitioning matrix $\mathbf{K}_{\textup{SF}}$ and lifting matrix $\mathbf{T}_{\textup{SF}}$ as follows. If $\mathbf{K}^*(i,j)<m_{\textup{f},d}+m_{\textup{n},d}+1$, then $\mathbf{K}_{\textup{SF}}(i,j)=\mathbf{K}^*(i,j)$; otherwise, $\mathbf{K}_{\textup{SF}}(i,j)=-1$. The same idea is applied for the lifting matrix. To obtain the SF-SC code with memory $m_{\textup{n},0}$, we do the following. If $\mathbf{K}^*(i,j)<m_{\textup{n},0}+1$, then $\mathbf{K}_{\textup{SF}}(i,j)=\mathbf{K}^*(i,j)$; otherwise, $\mathbf{K}_{\textup{SF}}(i,j)=-1$. The same idea is applied for the lifting matrix. Because of the way partitioning and lifting matrices are obtained, one can notice that protograph base matrices of corresponding RMC-SC and SF-SC codes are identical. Consequently, SF-SC codes have the same VN degrees as those of corresponding RMC-SC codes, which guarantees a fair comparison.

We have two groups of parameters for the codes we simulated. The first group is $(\gamma, \kappa, z, L, m_{\textup{n},0}, m_{\textup{n},1}, m_{\textup{n},2})=(7, 23, 23, 12, 6, 2, 3)$, and the second is $(\gamma, \kappa, z, L, m_{\textup{n},0}, m_{\textup{n},1}, \allowbreak m_{\textup{n},2})=(7, 35, 29, 16, 8, 3, 4)$. Since we have three memory values defined for each group, each group has three RMC-SC codes and three SF-SC codes. From now on, we will use the notation RMC-SC Code~$x.y$ to refer to the $x$'th group of parameters and the $y$'th RMC-SC code ($y=d$). For example, RMC-SC Code~1.0 refers to the first group and the RMC-SC code with memory $m_{\textup{n},0}=6$. RMC-SC Code~2.1 refers to the second group and the RMC-SC code with memory $m_{\textup{f},1}+m_{\textup{n},1}=11$. The notation SF-SC Code~$x.y$ is also adopted for SF-SC codes. RMC-SC as well as SF-SC Codes~1.0, 1.1, and 1.2 have design rates $0.5435$, $0.4928$, and $0.4167$, respectively. The code length for this group of parameters is $6{,}348$. RMC-SC as well as SF-SC Codes~2.0, 2.1, and 2.2 have design rates $0.7000$, $0.6625$, and $0.6125$, respectively. The code length for this group of parameters is $16{,}240$.

\vspace{-0.7em}
\begin{table}[H] 
\caption{Statistics of the Number of Cycles in RMC-SC Codes Compared With SF-SC Codes}
\centering
\scalebox{0.98}{
\begin{tabular}{|c|c|c|c|c|c|}
\hline
\makecell{Code} & \makecell{Cycle-$6$ \\ count} & \makecell{Cycle-$8$ \\ count} & Code & \makecell{Cycle-$6$ \\ count} & \makecell{Cycle-$8$ \\ count} \\ \hline
\begin{tabular}[c]{@{}c@{}}SF-SC \\ Code 1.0\end{tabular} & $2{,}047$     & $124{,}660$   & \begin{tabular}[c]{@{}c@{}}SF-SC\\ Code 2.0\end{tabular}  & $9{,}512$     & $748{,}664$   \\ \hline
\begin{tabular}[c]{@{}c@{}}RMC-SC\\ Code 1.0\end{tabular} & $0$        & $46{,}437$    & \begin{tabular}[c]{@{}c@{}}RMC-SC\\ Code 2.0\end{tabular} & $0$        & $319{,}493$   \\ \hline
\begin{tabular}[c]{@{}c@{}}SF-SC\\ Code 1.1\end{tabular}  & $4{,}623$     & $370{,}392$   & \begin{tabular}[c]{@{}c@{}}SF-SC\\ Code 2.1\end{tabular}  & $20{,}619$    & $2{,}068{,}947$  \\ \hline
\begin{tabular}[c]{@{}c@{}}RMC-SC\\ Code 1.1\end{tabular} & $184$      & $292{,}859$   & \begin{tabular}[c]{@{}c@{}}RMC-SC\\ Code 2.1\end{tabular} & $1{,}392$     & $1{,}459{,}976$  \\ \hline
\begin{tabular}[c]{@{}c@{}}SF-SC\\ Code 1.2\end{tabular}  & $25{,}530$    &          & \begin{tabular}[c]{@{}c@{}}SF-SC\\ Code 2.2\end{tabular}  & $120{,}727$   &          \\ \hline
\begin{tabular}[c]{@{}c@{}}RMC-SC\\ Code 1.2\end{tabular} & $31{,}763$    &          & \begin{tabular}[c]{@{}c@{}}RMC-SC\\ Code 2.2\end{tabular} & $153{,}845$   &          \\ \hline
\end{tabular}
}
\label{table_numeric}
\end{table}

In Table~\ref{table_numeric}, we compare cycle-$6$ and cycle-$8$ counts in RMC-SC codes and SF-SC codes for the two parameter groups we introduced above. Table~\ref{table_numeric} shows that RMC-SC codes offer significant reductions in the number of cycles-$6$ and cycles-$8$ relative to their SF-SC counterparts. For the first group, RMC-SC Code~1.0 completely eliminates cycles-$6$, whereas SF-SC Code~1.0 has $2{,}047$ cycles-$6$, i.e., $100\%$ reduction. RMC-SC Code~1.0 also achieves a $62.75\%$ reduction in the number of cycles-$8$ compared with SF-SC Code~1.0. RMC-SC Code~1.1 achieves reductions of $96.02\%$ in cycle-$6$ and $20.93\%$ in cycle-$8$ counts relative to SF-SC Code~1.1. SF-SC Code~1.2 has $19.62\%$ less cycles-$6$ than RMC-SC Code~1.2 because SF-SC Code~1.2 is constructed directly from the locally-optimal distribution given its memory. For the second group of parameters, RMC-SC Code~2.0 eliminates all cycles-$6$, while SF-SC Code~2.0 has $9{,}512$ cycles-$6$, i.e., again $100\%$ reduction. RMC-SC Code~2.0 achieves a $57.32\%$ reduction in the number of cycles-$8$ compared with SF-SC Code~2.0. RMC-SC Code~2.1 achieves reductions of $93.25\%$ in cycle-$6$ and $29.43\%$ in cycle-$8$ counts relative to SF-SC Code~2.1. SF-SC Code~2.2 has $21.53\%$ less cycles-$6$ than RMC-SC Code~2.2 for the same aforementioned reason.

\begin{figure}
\center
\hspace{-0.3em}\includegraphics[trim={0.5in 0.7in 0.6in 0.9in}, width=3.5in]{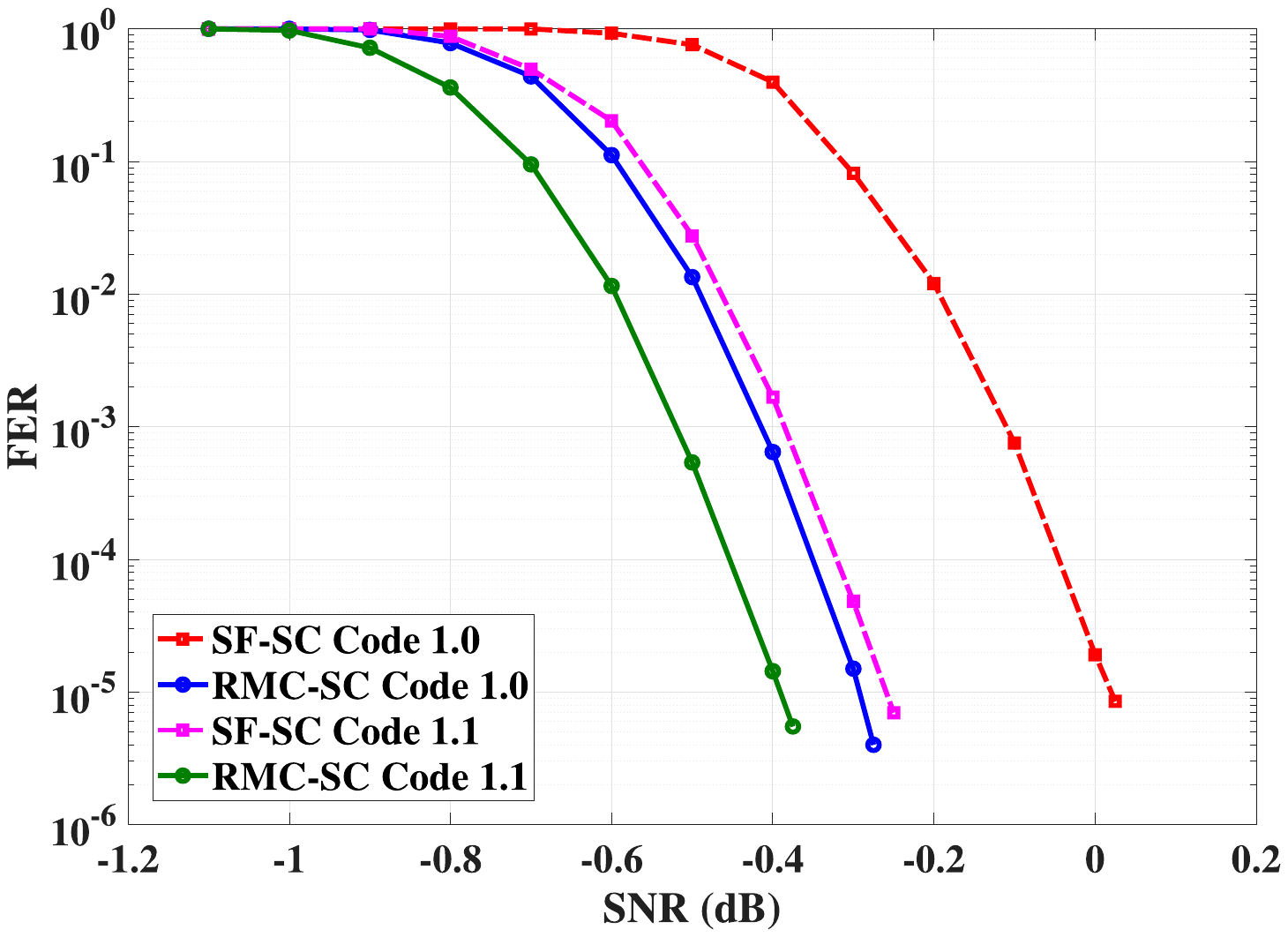}
\vspace{-1.5em}
\caption{FER versus SNR curves of RMC-SC and SF-SC codes designed with the first group of parameters, $(\gamma, \kappa, z, L, m_{\textup{n},0}, m_{\textup{n},1}, m_{\textup{n},2})=(7, 23, 23, \allowbreak 12, 6, 2, 3)$, over the AWGNC.}
\label{fig:perf1}
\vspace{-1.3em}
\end{figure}

Our simulations were conducted using a sum-product decoder that is based on fast Fourier transform. Moreover, for the additive white Gaussian noise channel (AWGNC), the signal-to-noise ratio (SNR) we adopt with the bipolar signal is $E_\textup{c}/N_0$, which is the energy per coded bit to noise power spectral density ratio. Such SNR definition is more convenient for applications where code reconfiguration is performed. In particular, incurring a rate loss to achieve a performance gain, via increasing the SC code memory here, is naturally acceptable in such applications.

In Fig.~\ref{fig:perf1}, we show the frame error rate (FER) performance of RMC-SC and SF-SC codes designed with the first parameter group over the AWGNC. The results demonstrate remarkable performance gains achieved via our RMC-SC codes with respect to SF-SC codes. In particular, at SNR $-0.3$ dB, RMC-SC Code~1.0 achieves an FER performance gain of about $3.73$ orders of magnitude compared with SF-SC Code~1.0. RMC-SC Code~1.0 also offers a gain of about $0.31$ dB relative to SF-SC Code~1.0 at FER $\approx1.50\times10^{-5}$. Moreover, at SNR $-0.4$ dB, RMC-SC Code~1.1 achieves an FER performance gain of about $2.07$ orders of magnitude compared with SF-SC Code~1.1. RMC-SC Code~1.1 also offers a gain of about $0.13$ dB relative to SF-SC Code~1.1 at FER $\approx1.44\times10^{-5}$.

\begin{figure}
\center
\hspace{-0.3em}\includegraphics[trim={0.5in 0.7in 0.6in 0.9in}, width=3.5in]{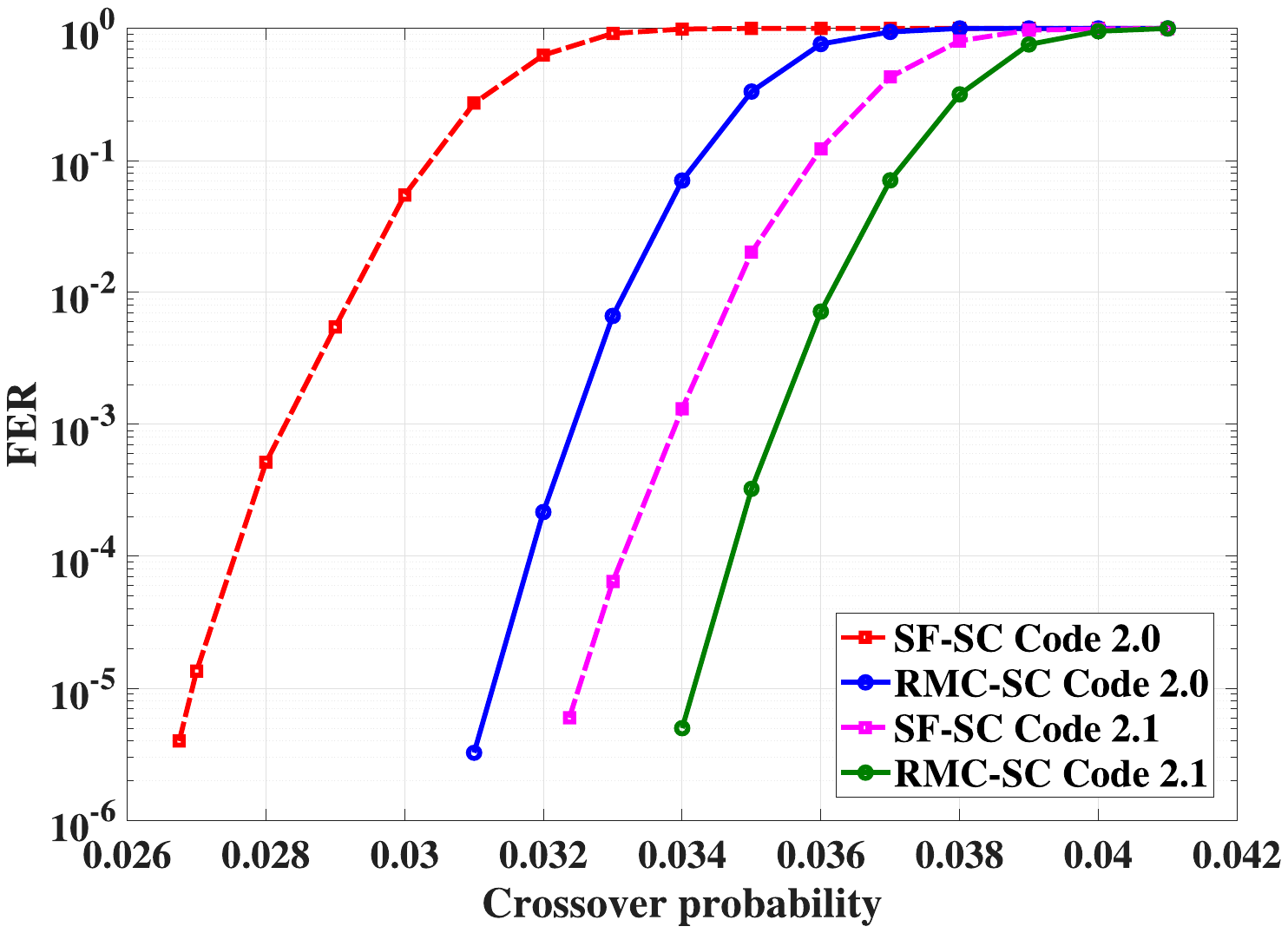}
\vspace{-1.5em}
\caption{FER versus crossover probability curves of RMC-SC and SF-SC codes designed with the second group of parameters, $(\gamma, \kappa, z, L, m_{\textup{n},0}, m_{\textup{n},1}, \allowbreak m_{\textup{n},2})=(7, 35, 29, 16, 8, 3, 4)$, over the BSC.}
\label{fig:perf2}
\vspace{-1.2em}
\end{figure}

In Fig.~\ref{fig:perf2}, we present the FER performance of RMC-SC and SF-SC codes designed with the second parameter group (longer codes) over the binary symmetric channel (BSC). The results validate the remarkable performance gains RMC-SC codes offer with respect to SF-SC codes. At crossover probability $0.031$, RMC-SC Code~2.0 achieves an FER performance gain of about $4.92$ orders of magnitude compared with SF-SC Code~2.0. Furthermore, at crossover probability $0.034$, RMC-SC Code~2.1 achieves an FER performance gain of about $2.42$ orders of magnitude compared with the SF-SC Code~2.1.

Unlike SF-SC codes, RMC-SC codes are optimized at~each design stage, which justifies the significant cycle-count reductions and remarkable performance gains they offer. When the memory is $m_s$, RMC-SC Codes~1.2 and 2.2 are quite close to SF-SC Codes~1.2 and 2.2, respectively, in terms of performance. Intriguingly, RMC-SC codes may slightly outperform SF-SC codes even in this case.

We suggest that an additional reason behind the performance gains offered by our RMC-SC codes is their CN degree distributions as implied by Fig.~\ref{fig:gd_dist}. To demonstrate that, we performed first-order threshold analysis based on VN and CN degree distributions for RMC-SC and SF-SC codes with the second group of parameters in two different settings \cite{urbanke_de,bazzi_de}. The first setting has the BSC along with the Gallager A decoder, while the second setting has the binary erasure channel (BEC) along with the peeling decoder. As shown in Table~\ref{table_de}, density evolution analysis reveals up to $14.55\%$ threshold gain in the first setting and up to $7.31\%$ threshold gain in the second setting achieved by our RMC-SC codes over SF-SC codes. These gains are consistent with those reported via the FER performance plots in Fig.~\ref{fig:perf1} and Fig.~\ref{fig:perf2}, but also with different channels or/and decoders.

Lastly, we evaluate the hardware savings that result from using our RMC-SC codes compared with using three distinct SC codes with no reconfiguration. The encoding-decoding complexity is dictated by the number of edges in the graph representation of the code. The comparison is against SC codes that have the same total number of edges in their graphs as our RMC-SC codes. Conveniently, our computations here are based on the number of $1$'s in the protograph base matrices. Let that number be $T_{x.y}$ for RMC-SC Code~$x.y$. Then, the hardware reduction, percentage-wise, is given by:
\begin{align}
&\textup{Hardware reduction} = \nonumber \\ &\frac{T_{x.0}+(T_{x.0}+T_{x.1})}{T_{x.0}+(T_{x.0}+T_{x.1})+(T_{x.0}+T_{x.1}+T_{x.2})} \cdot 100\%.
\end{align}
For the first parameter group ($x=1$), the hardware reduction is $55.28\%$. For the second parameter group ($x=2$), the hardware reduction is $54.88\%$.

\begin{table}
\caption{Thresholds via Density Evolution Analysis (Crossover Probabilities for BSC and Erasure Probabilities for BEC)}
\centering
\scalebox{0.98}{
\begin{tabular}{|c|c|c|c|c|}
\hline
\makecell{Setting} & \makecell{Code} & \makecell{SF-SC} & \makecell{RMC-SC} & \makecell{Gain (\%)} \\ \hline
\multirow{2}{*}{\makecell{BSC \\ Gallager-A}} 
& Code 2.0  & $0.0110$ & $0.0126$ & $14.55\%$ \\ \cline{2-5} 
& Code 2.1  & $0.0160$ & $0.0167$ & $4.38\%$  \\ \hline
\multirow{2}{*}{\makecell{BEC \\ Peeling}}    
& Code 2.0  & $0.2190$ & $0.2350$ & $7.31\%$  \\ \cline{2-5} 
& Code 2.1  & $0.2340$ & $0.2424$ & $3.59\%$  \\ \hline
\end{tabular}
}
\label{table_de}
\end{table}

\vspace{-0.2em}
\section{Conclusion}\label{sec_conc}

We proposed a new class of RC-SC codes, which we named RMC-SC codes, where rate compatibility is achieved via increasing the code memory. We introduced a systematic recursive framework to design RMC-SC codes probabilistically in consecutive design stages, aiming to minimize the number of short cycles. We expressed the expected number of detrimental cycles in the RMC-SC code protograph in terms of probability distributions that characterize the partitioning. We developed a gradient-descent algorithm that finds a locally-optimal distribution for the new components at each design stage to guide the FL design. We customized an MCMC algorithm to then perform the FL optimization at the partitioning and lifting phases to design these new components. Experimental results reveal up to $100\%$ reduction in certain cycle counts as well as performance gains in orders of magnitude achieved by RMC-SC codes compared with SF-SC codes. Our method can be extended to other rate-compatibility approaches. Future work includes combining RMC-SC codes with machine learning for adaptive coding in practical systems in addition to extended threshold analysis of these codes. We suggest that RMC-SC codes can be a valuable tool to remarkably enhance reliability in modern communication and storage systems.

\section*{Acknowledgment}\label{sec_ack}

This work was supported in part by the T\"{U}B\.ITAK 2232-B International Fellowship for Early Stage Researchers.

\end{document}